\documentclass[a4paper,12pt]{article}
\usepackage{amsfonts}
\usepackage{latexsym}
\usepackage{amsmath}
\usepackage{amssymb}
\usepackage{amssymb}
\usepackage{slashed}
\usepackage{upgreek}
\usepackage{mathrsfs}
\usepackage{bbm}

\usepackage{relsize}
\usepackage{graphicx}
\usepackage{comment}

\newcommand{\newsection}{    
\setcounter{equation}{0}\section}

 \usepackage[title]{appendix}
  \renewcommand{\appendix}[1]{\section{#1}}

\def\bbe{{\bf{e}}}
\def\bbl{{\bf{\ell}}}
\font\mybb=msbm10 at 11pt

\def\bb#1{\hbox{\mybb#1}}

\def\bZ {\bb{Z}}
\def\bR {\bb{R}}

\def\gX{\Gamma\mkern-4.0mu X}
\def\gom{\Gamma\mkern-4.0mu \omega}
\def\gY{\Gamma\mkern-2.0mu Y}

\def\gQ{\Gamma\mkern-4.0mu Q}
\def\gH{\Gamma\mkern-4.0mu H}

\def\sX{\slashed {X}}
\def\sgX{\slashed {\gX}}
\def\sY{\slashed {Y}}
\def\sgY{\slashed {\gY}}

\def\sF{\slashed {F}}

\def\sQ{\slashed {Q}}
\def\sgQ{\slashed {\gQ}}

\def\sH{\slashed{H}}
\def\sgH{\slashed{\gH}}

\newcommand{\bea}{\begin{eqnarray}}
\newcommand{\eea}{\end{eqnarray}}

\usepackage[usenames,dvipsnames,svgnames,table]{xcolor}	
\usepackage[normalem]{ulem}				
\usepackage{microtype}
\usepackage[colorlinks, citecolor=blue, linkcolor=blue, urlcolor=Maroon, filecolor=Maroon, linktocpage=true]{hyperref} 

\begin{document}

\begin{center}
\vspace*{-1.0cm}
\begin{flushright}
\normalsize{\texttt{ZMP-HH/18-10}}\\
\end{flushright}


\vspace{2.0cm} {\Large \bf A non-existence theorem for $N>16$ supersymmetric  AdS$_3$ backgrounds  } \\[.2cm]

\vskip 2cm
A.~S.~Haupt$^{1,2}$, S.  Lautz$^3$ and  G.  Papadopoulos$^3$
\\
\vskip .6cm

\begin{small}
$^1$\textit{Department of Mathematics \\ and Center for Mathematical Physics,\\
University of Hamburg,\\ Bundesstr. 55, D-20146 Hamburg, Germany}
\end{small}\\*[.6cm]

\begin{small}
$^2$\textit{II. Institute for Theoretical Physics,\\ University of Hamburg,\\
Luruper Chaussee 149, D-22761 Hamburg, Germany}\\
\texttt{alexander.haupt@uni-hamburg.de}
\end{small}\\*[.6cm]

\begin{small}
$^3$\textit{Department of Mathematics, King's College London
\\
Strand, London WC2R 2LS, UK}\\
\texttt{sebastian.lautz@kcl.ac.uk}\\
\texttt{  george.papadopoulos@kcl.ac.uk}
\end{small}\\*[.6cm]

\end{center}

\vskip 2.5 cm

\begin{abstract}
\noindent
We show that there are no smooth warped   AdS$_3$ solutions in  10- and 11-dimensional supergravities which preserve strictly more than 16 supersymmetries
and have  internal space a compact without boundary manifold.
	
\end{abstract}

\newpage


\section{Introduction}

The main purpose of this paper is to complete the classification of AdS backgrounds that preserve strictly more than 16 supersymmetries in 10- and 11-dimensional supergravities.
Such backgrounds have found many applications originally in supergravity compactifications and more recently in the AdS/CFT correspondence, for reviews see \cite{duff, maldacena}.
  In the latter case    AdS backgrounds that preserve $N>16$ supersymmetries  are associated with
the best understood examples of the  correspondence \cite{maldacenab, abjm}.

The maximally supersymmetric AdS backgrounds\footnote{The maximally supersymmetric AdS backgrounds are of the type
considered previously in \cite{FR}.}   have been classified  in \cite{maxsusy} and it has been found that they are locally isometric
to the  $\text{AdS}_4\times S^7$ \cite{duffpopemax} and  $\text{AdS}_7\times S^4$ \cite{townsend} solutions of  11-dimensional supergravity,  and to the $\text{AdS}_5\times S^5$ solution of  IIB supergravity, see \cite{schwarz} and comment within. There are no  $\text{AdS}_7$ backgrounds
that preserve $16<N<32$ supersymmetries \cite{mads, iibads, iiaads} and no smooth $\text{AdS}_6$ backgrounds that preserve $N>16$ supersymmetries with compact without boundary  internal space \cite{ads6}.  More recently, it has been demonstrated under the same assumptions on the internal space that there are
no smooth  $\text{AdS}_5$ backgrounds  that preserve $16<N<32$ \cite{ads5};  see \cite{ferrara, aharony, garcia} for applications to AdS/CFT.  It has also been shown  in \cite{ads4Ngr16} that the only smooth AdS$_4$ solution
with compact without boundary internal space that preserves $16<N<32$ supersymmetries is locally isometric to the $N=24$ IIA solution $\text{AdS}_4\times\mathbb{CP}^3$ of \cite{nilssonpope}.  Moreover
it has been shown in \cite{ads2} that there are no smooth AdS$_2$ backgrounds in 10- and 11-dimensional supergravities with compact without boundary internal space that preserve $N>16$
supersymmetries.  Product solutions $\text{AdS}_n\times M^{D-n}$  with $M^{D-n}$ a symmetric space have been classified   in \cite{figueroaa}-\cite{Wulffb}. Furthermore
the geometry of all heterotic AdS$_3$ backgrounds have been investigated  in \cite{hetads} and it has been found that there are no solutions that preserve  $N>8$ supersymmetries.

The only class of backgrounds that remains to be investigated are the warped AdS$_3$ backgrounds with the most general allowed fluxes in 10-dimensional type II and 11-dimensional supergravity theories
that preserve $N>16$ supersymmetries. For these,   we shall demonstrate a non-existence theorem provided that the solutions are smooth
and their internal spaces are  compact manifolds without boundary. It suffices to establish the non-existence theorem up to  local isometries.  The more general
result follows as there are no new geometries that can be constructed by taking quotients by discrete groups.

The method used to establish this result relies on a number of recent developments.  One of them is the integration of both field equations and Killing spinor equations (KSEs)
of 10- and 11-dimensional supergravities over the AdS$_n$ subspace for all warped    AdS backgrounds, $\text{AdS}_n\times_w M^{D-n}$, without making any other assumptions
on the fields and Killing spinors apart from imposing the symmetries of the AdS subspace on the fields \cite{mads, iibads, iiaads}. This integration reduces the field equations and KSEs of 10- and 11-dimensional  supergravity theories to a system of independent equations  on the internal space
$M^{D-n}$. To solve these, we use  another key development which is the homogeneity theorem in \cite{homogen}.  This states that all 10- and 11-dimensional supergravity
backgrounds that preserve $N>16$ supersymmetries are Lorentzian homogeneous spaces and all the fields are invariant tensors. Another ingredient for the proof   is the classification of Killing superalgebras of all warped  AdS backgrounds in \cite{superalgebra}. For AdS backgrounds that preserve $N>16$ supersymmetries this enables us to identify all the Lorentzian algebras that act transitively and (almost) effectively\footnote{A group $G$ acts almost effectively on a space $M$, iff the group action induces an inclusion
 of $\mathfrak{Lie}\,G$ into the space of vector fields on $M$, see also section 2.  We also say that  $\mathfrak{Lie}\,G$  acts effectively on $M$.} on the spacetime. The requirement for a transitive action is a consequence of the homogeneity theorem while that of the (almost) effective
action is needed for the super-Jacobi identities of the superalgebra to be satisfied.

The strategy of the proof is as follows.  First one establishes that for all $\text{AdS}_3\times_w M^{D-3}$ backgrounds that preserve $N>16$ supersymmetry, the warp factor is constant. Therefore
 the geometry is a product $\text{AdS}_3\times M^{D-3}$. To show this, either
one uses that the solutions are smooth and the internal space is compact without boundary as well as techniques from the proof of the homogeneity theorem or that the even subalgebra
$\mathfrak{g}_0$ of the Killing
superalgebra $\mathfrak{g}$ of $\text{AdS}_3\times_w M^{D-3}$ decomposes as $\mathfrak{g}_0=\mathfrak{iso}(\text{AdS}_3)\oplus \mathfrak{iso} (M^{D-3})$, where $\mathfrak{iso}(AdS_3)$
is  an isometry algebra of AdS$_3$
and   $\mathfrak{iso} (M^{D-3})=\mathfrak{t}_0$  is an isometry algebra of the internal space $M^{D-3}$.

Having established that the $N>16$ AdS$_3$ backgrounds are products, $\text{AdS}_3\times M^{D-3}$ and that $\mathfrak{g}_0=\mathfrak{iso}(\text{AdS}_3)\oplus \mathfrak{t}_0$,
where $\mathfrak{t}_0$ is an algebra of isometries on $M^{D-3}$, we obtain as a consequence of the homogeneity theorem that the internal space is a homogeneous space $G/H$
with $\mathfrak{Lie} \,G=\mathfrak{t}_0$.  In addition, the theorem requires that all  fields are invariant under the left action of $G$ on $G/H$.

The final part of the proof involves the identification of all homogeneous spaces\footnote{As we are investigating supersymmetric backgrounds, we require that all the internal spaces are spin.}  in seven and eight dimensions that admit a transitive and an almost effective  action of a group $G$
with Lie algebra $\mathfrak{t}_0$.  For $\mathfrak{t}_0$  semisimple, one   can identify the relevant  homogeneous spaces     using  the classification results of (simply connected)  7- and 8-dimensional homogeneous
manifolds in  \cite{castellani}-\cite{bohmkerr}; for a concise description see \cite{klausthesis}.  There is also the need of a  procedure on homogeneous spaces  which  tests whether a $\mathfrak{t}_0$ can act effectively on a given $G/H$.
 In section 2,  we refer to it as ``modification'' of a homogeneous space. A similar approach can be used for the case that
 $\mathfrak{t}_0$  is not semisimple.   After identifying all the suitable homogeneous spaces,  a substitution of the geometric data into the field equations and KSEs of supergravity
theories in 10- and 11-dimensions establishes our non-existence theorem.

Before we proceed with the proof, let us investigate the need for the assumptions we have made. First
 one can establish that if  $\text{AdS}_3\times_w M^{D-3}$ is smooth and $M^{D-3}$   is compact without boundary, then the even subalgebra of the Killing superalgebra
of  $\text{AdS}_3\times_w M^{D-3}$ will decompose as $\mathfrak{g}_0=\mathfrak{iso}(\text{AdS}_3)\oplus \mathfrak{t}_0$ \cite{superalgebra}.  The requirement that $M^{D-3}$   must be compact without boundary
may be weakened but not completely removed. If it is removed, then $\mathfrak{g}_0$ may  not decompose as stated above. In addition the warp factor of $\text{AdS}_3\times_w M^{D-3}$
backgrounds with $N>16$ supersymmetries may not be constant and  there exist AdS$_3$ backgrounds that preserve $N>16$ supersymmetries, see \cite{dsads} for a detailed exposition.  In particular the maximally supersymmetric $\text{AdS}_4\times S^7$ and  $\text{AdS}_7\times S^4$ backgrounds of 11-dimensional supergravity can be
viewed as warped AdS$_3$ backgrounds but the internal spaces are not compact without boundary.

The paper is organized as follows. In section 2, we describe the Killing superalgebras of AdS$_3$ backgrounds and introduce the notion of a modification
of a homogeneous space which allows us to test whether an algebra can act effectively on it.  In sections 3, 4 and 5, we prove the main result of our paper
for 11-dimensional, IIA and IIB supergravities, respectively. In appendix A, we describe our conventions.  In appendix B, we present some aspects of the structure
of homogeneous spaces admitting a transitive action by a compact but not semisimple Lie group that are useful in the proof of our results,   and in appendix C,
we describe the geometry of the $N^{k,l}$ homogeneous space.

\section{Symmetries of AdS$_3$ backgrounds}\label{superxx}

\subsection{ Killing superalgebras of AdS$_3$ backgrounds }

As AdS$_3$ is locally a group manifold, the
Killing superalgebras of warped AdS$_3$ backgrounds with the most general allowed fluxes  decompose as $\mathfrak{g}=\mathfrak{g}_L\oplus \mathfrak{g}_R$, where $\mathfrak{g}_L$ and  $\mathfrak{g}_R$  are associated with the
left and right actions. The left and right Killing superalgebras $\mathfrak{g}_L$ and $\mathfrak{g}_R$ have been identified in \cite{superalgebra}.
This has been done under the assumptions that either the internal space is compact without boundary or that the even  subalgebra decomposes as
$(\mathfrak{g}_L{})_0=\mathfrak{sl}(2, \bR)_L\oplus (\mathfrak{t}_L)_0$ and similarly for $ \mathfrak{g}_R$.  As $\mathfrak{g}_0=\mathfrak{iso}(\text{AdS}_3)\oplus \mathfrak{t}_0=(\mathfrak{g}_L{})_0\oplus (\mathfrak{g}_R{})_0$,
$\mathfrak{iso}(\text{AdS}_3)$
is isomorphic to either $\mathfrak{sl}(2, \bR)_L$ or $\mathfrak{sl}(2, \bR)_R$ if the background has only either left or right supersymmetries, respectively, or $\mathfrak{iso}(\text{AdS}_3)=\mathfrak{sl}(2, \bR)_L\oplus \mathfrak{sl}(2, \bR)_R$ if the background has both left and right supersymmetries. Furthermore $\mathfrak{t}_0= (\mathfrak{t}_L)_0\oplus (\mathfrak{t}_R)_0$.

 It has been shown in \cite{superalgebra} that for AdS$_3$ backgrounds $\mathfrak{t}_0$ may not be semisimple and in addition may admit   central terms $\mathfrak{c}$ which commute
 with all other generators of the superalgebra.  We shall show below that in all cases but one $\mathfrak{c}_L=\{0\}$.  If $\mathfrak{c}_L\not=\{0\}$,  it will have at most dimension 3.  The left and right superalgebras are isomorphic
 and so it suffices to present only the left ones. These are tabulated in table\footnote{Throughout this paper $\mathfrak{sp}(n)$, $n\geq1$, denotes the compact symplectic
 Lie algebras.  These have been denoted  with $\mathfrak{sp}^*(n)$ in \cite{superalgebra} to distinguish them from the non-compact ones.}  \ref{table:ads3ksa}.

 \subsection{Central terms}

We shall focus on $\mathfrak{g}_L$ as the description that follows  below also applies to $\mathfrak{g}_R$.
It has been observed in \cite{superalgebra} that the Killing superalgebras of AdS$_3$ backgrounds may exhibit central terms.  Such terms may occur in all cases apart from $\mathfrak{osp}(n|2)$ and
$\mathfrak{D}(2,1,\alpha)$.
However it has been shown in \cite{superalgebra} that both $\mathfrak{f}(4)$ and $\mathfrak{g}(3)$  exhibit such terms.  Though $\mathfrak{sl}(2|2)/1_{4\times 4}$ can exhibit up to three central terms.
This is because $\mathfrak{sl}(2|2)/1_{4\times 4}$ arises as a special case of  $\mathfrak{D}(2,1,\alpha)$ at special values of the parameter $\alpha$.  At those values three of the R-symmetry
generators of $\mathfrak{D}(2,1,\alpha)$ span  the R-symmetry algebra  $\mathfrak{so}(3)$ of  $\mathfrak{sl}(2|2)/1_{4\times 4}$ and the other three become central.

It can also be shown that $\mathfrak{sl}(n|2)$, $n>2$ and $\mathfrak{osp}(4|2n)$, $n>1$,  do not exhibit central terms either.  This can be seen after an analysis of the
condition
\bea
\alpha_{rsr'}{}^t \tilde V_{ts'}-
\alpha_{rss'}{}^t \tilde V_{tr'}+\alpha_{r's'r}{}^t \tilde V_{ts}-
\alpha_{r's's}{}^t \tilde V_{tr}=0~,
\label{consis}
\eea
of \cite{superalgebra}, where $\tilde V_{rs}=-\tilde V_{sr}$ are the generators of $(\mathfrak{t}_L)_0$ and $\alpha$ is described in \cite{superalgebra}.  For $\mathfrak{sl}(n|2)$, $n>2$, the central terms that can occur are  (2,0) and (0,2) components of the $\tilde V$.  However one can show that these do not satisfy (\ref{consis}) unless they vanish.  Thus $\mathfrak{c}=\{0\}$.

 It remains to investigate the superalgebra with $(\mathfrak{t}_L)_0/\mathfrak{c}_L= \mathfrak{sp}(n) \oplus \mathfrak{sp}(1)$. The central generators that can occur are the $\tilde V$
 which lie in the complement of $\mathfrak{sp}(n) \oplus \mathfrak{sp}(1)$ in $\mathfrak{so}(4n)$.
Taking the trace of (\ref{consis}) with one of the three complex structures that are associated with $\mathfrak{sp}(n) \oplus \mathfrak{sp}(1)$, one can  demonstrate that all such generators  $\tilde V$ must also vanish.  Thus again $\mathfrak{c}=\{0\}$. Therefore apart from  $\mathfrak{sl}(2|2)/1_{4\times 4}$, all the other superalgebras in  table \ref{table:ads3ksa} do not exhibit central terms.

 \begin{table}
	\caption{$AdS_3$ Killing superalgebras in type II and 11D}
	\centering
	\begin{tabular}{|c|c|c|c|}
		\hline
		$N_L$ & $\mathfrak{g}_L/\mathfrak{c}_L$ & $(\mathfrak{t}_L)_0/\mathfrak{c}_L$&$\mathrm{dim}\, \mathfrak{c}_L $ \\
		\hline
		$2n$& $\mathfrak{osp}(n|2)$ & $\mathfrak{so}(n)$ & 0\\
		$4n,~n>2$ & $\mathfrak{sl}(n|2)$ & $\mathfrak{u}(n)$& 0 \\
		$8n, n>1$ & $\mathfrak{osp}(4|2n)$ & $\mathfrak{sp}(n) \oplus \mathfrak{sp}(1)$ &0\\
		16 & $\mathfrak{f}(4)$ & $\mathfrak{spin}(7)$ &0\\
		14 & $\mathfrak{g}(3)$ & $\mathfrak{g}_2$&0 \\
		8 & $\mathfrak{D}(2,1,\alpha)$ & $\mathfrak{so}(3) \oplus \mathfrak{so}(3)$&0 \\
		8 & $\mathfrak{sl}(2|2)/1_{4\times 4}$ & $\mathfrak{su}(2)$& $\leq 3 $\\ [1ex]
		\hline
	\end{tabular}
	\label{table:ads3ksa}
\end{table}

\subsection{On the $G/H$ structure of internal spaces} \label{modifx}

We shall demonstrate later that the spacetime of all AdS$_3$ backgrounds that preserve $>16$ supersymmetries in 10- and 11-dimensional supergravities  is a product $\text{AdS}_3\times M^{D-3}$
and that $M^{D-3}$ is a homogeneous space $M^{D-3}=G/H$ such that $\mathfrak{Lie} G=\mathfrak{t}_0$. Of course $G$ acts transitively on $M^{D-3}$.  In addition it is required to
 act  ``almost effectively'' on $M^{D-3}$.  This means that the map of $\mathfrak{Lie}\, G$ into the space of Killing vector fields of $M^{D-3}$ is an inclusion, ie for every generator of
 $\mathfrak{Lie}\, G$ there is an associated {\it non-vanishing} Killing vector field on $M^{D-3}$.   We shall also refer to this property as $\mathfrak{Lie}\, G$ acting ``effectively'' on $M^{D-3}$.
This latter property is essential as otherwise  the super-Jacobi identities of the AdS  Killing superalgebra  will not be satisfied. It is also essential for the identification of the
manifolds that can arise as internal spaces of all AdS, and in particular AdS$_3$,  backgrounds preserving some supersymmetry.

For AdS$_3$ backgrounds, there are two cases to consider.  The first case arises  whenever $\mathfrak{t}_0$ is a simple
Lie algebra.  Then the internal spaces can be identified, up a to factoring with a  finite group, using the classification of the simply connected 7- and 8-dimensional homogeneous spaces
 in  \cite{castellani}-\cite{bohmkerr}.  This is sufficient to identify the internal spaces of all such AdS$_3$ backgrounds that preserve $N>16$ supersymmetries.

However for most AdS$_3$ backgrounds $\mathfrak{t}_0$ is not simple.  Typically it is the sum of two Lie algebras, $\mathfrak{t}_0=(\mathfrak{t}_L)_0\oplus(\mathfrak{t}_R)_0$,  one arising from the left sector and another from the right sector. In addition, it may not be  semisimple.  For example, we have seen that $\mathfrak{t}_0=\mathfrak{u}(3)=\mathfrak{su}(3)\oplus \mathfrak{u}(1)$ for    the  $\mathfrak{sl}(3|2)$ Killing superalgebra  and    $\mathfrak{t}_0=\mathfrak{su}(2)\oplus \mathfrak{c}$ for the    $\mathfrak{sl}(2|2)/1_{4\times 4}$ superalgebra with a central term $\mathfrak{c}$.  Furthermore,  $\mathfrak{t}_0$ is not semisimple for all AdS$_3$ backgrounds that exhibit either $N_L=4$ or $N_R=4$ supersymmetries. Given that $\mathfrak{t}_0$ may not be simple, the question then arises how one can decide given a $G'/H'$ space chosen from the classification results of \cite{castellani}-\cite{bohmkerr} whether $\mathfrak{t}_0$ acts both
transitively and effectively on $G'/H'$.

Let us illustrate this with  examples.  It is known that both $U(n)$ and $SU(n)$ act transitively and effectively on $S^{2n-1}$.  Thus $S^{2n-1}=U(n)/U(n-1)$ and $S^{2n-1}=SU(n)/SU(n-1)$.
However for $n>2$, it is  $\mathfrak{u}(n)$ which appears as a subalgebra of $\mathfrak{sl}(n|2)$ and so $\mathfrak{u}(n)$ is expected to act transitively and effectively
on the internal spaces instead of $\mathfrak{su}(n)$.   From this perspective  $U(n)/U(n-1)$ can arise as a potential internal space of an AdS$_3$ background whereas $SU(n)/SU(n-1)$ should be discarded.  As in the classification results for homogeneous spaces it is not apparent which description is used for a given homogeneous  space but essential for the classification of AdS$_3$ backgrounds, let us investigate the above paradigm further. To see how $S^{2n-1}=SU(n)/SU(n-1)$ can be modified to be written as a $U(n)/U(n-1)$, consider the group homomorphism $i$ from $SU(n-1)\times U(1)$
into $SU(n)$ as
\bea
(A, z)\xrightarrow[]{i} \begin{pmatrix} A z &0\cr 0 & z^{1-n}\end{pmatrix}~.
\eea
In fact $i$ has kernel $\bZ_{n-1}$ and so factors to $U(n-1)$.
Next consider $SU(n)\times U(1)$ and the group homomorphism $j$ of  $SU(n-1)\times U(1)$ into $SU(n)\times U(1)$ as
\bea
(A, z)\xrightarrow{j} \left(\begin{pmatrix} A z &0\cr 0 &z^{1-n}\end{pmatrix}, z^{n-1}\right)~.
\eea
Again $j$  has kernel $\bZ_{n-1}$ and so factors to $U(n-1)$.
Then $SU(n)\times U(1)/j(SU(n-1)\times U(1))=S^{2n-1}$ with $SU(n)\times U(1)$ acting almost effectively on $S^{2n-1}$. Furthermore  one can verify that $U(n)=(SU(n)\times U(1))/\bZ_n$
acts effectively on $S^{2n-1}$ as expected.

The key point of the modification described above is the existence  of $U(1)\subset SU(n)$ such that  $SU(n-1)\times U(1)\subset SU(n)$ and that this $U(1)$ acts on both $SU(n)$ and
the $U(1)$ subgroup of $SU(n)\times U(1)$.  Observe that after the modification the isotropy group is larger and so the invariant geometry of $S^{2n-1}$ as a $U(n)/U(n-1)$
homogeneous space is more restrictive  than that of $S^{2n-1}=SU(n)/SU(n-1)$.

Another example that illustrates a similar point and which  will be used in the analysis that follows is $S^7=Sp(2)/Sp(1)$.
It is known that $S^7$ can also be described as $S^7=Sp(2)\cdot Sp(1)/Sp(1)\cdot Sp(1)$, where $Sp(2)\cdot Sp(1)=Sp(2)\times Sp(1)/\bZ_2$ and similarly for $Sp(1)\cdot Sp(1)$. The modification required to describe
$S^7$ as an $Sp(2)\cdot Sp(1)/Sp(1)\cdot Sp(1)$ coset starting from $Sp(2)/Sp(1)$ is as follows. View the elements of $Sp(2)$ as  $2\times 2$ matrices
with  quaternionic entries and consider the inclusion $i$ of $Sp(1)\times Sp(1)$ in $Sp(2)$ as
\bea
(x, y)\xrightarrow[]{i} \begin{pmatrix} x &0\cr 0 & y\end{pmatrix}~,
\eea
where $x$ and $y$ are quaternions of length one.
Then the map $j$ from $Sp(1)\times Sp(1)$ into $Sp(2)\times Sp(1)$ is constructed as
\bea
(x, y)\xrightarrow{j} \left(\begin{pmatrix} x &0\cr 0 &y\end{pmatrix}, y\right)~.
\label{modifsp2}
\eea
One finds that $Sp(2)\times Sp(1)/j(Sp(1)\times Sp(1))$ is diffeomorphic to $S^7$, with $Sp(2)\times Sp(1)$ acting almost effectively and descending to an effective action for $Sp(2)\cdot Sp(1)$.  Again the additional $Sp(1)$ introduced in the isotropy group acts both on $Sp(2)$ and the additional
$Sp(1)$ introduced in the transitive group.  The geometry of the homogeneous space $S^7=Sp(2)\cdot Sp(1)/Sp(1)\cdot Sp(1)$ is more restrictive than that
of $S^7=Sp(2)/Sp(1)$.  In fact the former is a special case of the latter.  As a final example $SU(2)\times SU(2)/SU(2)$ can be seen as a modification of the homogeneous space $SU(2)/\{e\}$.  From now on we shall refer to  such  constructions as ``modifications'' of a homogeneous space.

On the level of Lie algebras the  modifications can be viewed as follows.  Suppose $\mathfrak{t}_0$ decomposes as $\mathfrak{t}_0=\mathfrak{k}\oplus \mathfrak{e}$, where $\mathfrak{k}$ and $\mathfrak{e}$ are Lie algebras, and that
there is a homogeneous space $K/L$ with $\mathfrak{Lie} (K) =\mathfrak{k}$.  To see whether $K/L$ can be modified to admit an effective
action of the whole $\mathfrak{t}_0$ algebra, it is first required that $\mathfrak{l}\oplus\mathfrak{e}$ is a subalgebra of $\mathfrak{k}$, where $\mathfrak{Lie} L=\mathfrak{l}$.
Then, up to possible discrete identifications, $K/L$ can be modified to $K\times E/L\times E$, where now $E$ with $\mathfrak{Lie}\, E=\mathfrak{e}$ acts on both the $K$ and $E$ subgroups of the transitive group.

All    7- and 8-dimensional  $K/L$  homogeneous spaces with $K$ semisimple are known up to possible modifications.  Because of this for $\mathfrak{t}_0$ semisimple, one can systematically search for all  modifications to $K/L$ homogeneous spaces to find whether a Lie algebra $\mathfrak{t}_0$ can act transitively and
effectively on a modified homogenous space.  If $\mathfrak{t}_0$ is not semisimple, we have argued in appendix \ref{modif} that up to discrete identifications one can construct all the homogeneous spaces $G/H$ with $\mathfrak{Lie}\,G=\mathfrak{t}_0$ as product of a modification of a homogeneous space $K/L$ with $K$ semisimple with the abelian group $\times^k U(1)$.

As we shall see the modifications of homogeneous spaces are necessary  to identify all possible internal spaces of AdS$_3$ backgrounds that can preserve some supersymmetry.  For such modifications to exist for $K/L$
a necessary condition is that the rank of $L$ must be strictly smaller than that of $K$.  It turns out that this is rather restrictive in the analysis that follows.

Let us now turn to investigate the homogeneous geometry of a modification $K\times E/L\times E$ of the homogenous $K/L$ space. One can show that this can be explored as a special case of that  of $K/L$.  Indeed
suppose that $\mathfrak{k}=\mathfrak{l}\oplus \mathfrak{m}$.  Then observe that one can choose  the generators of $\mathfrak{Lie}(K\times E)$ such that
$\mathfrak{Lie}(K\times E)= j(\mathfrak{l}\oplus \mathfrak{e})\oplus \mathfrak{m}$, where $j:~~\mathfrak{l}\oplus \mathfrak{e}\rightarrow \mathfrak{k}\oplus \mathfrak{e}$ is the
inclusion of the modification.  Therefore the tangent space at the origin of the original $K/H$ space and that of the modification $K\times E/L\times E$ can be identified with the same vector space $\mathfrak{m}$.
The only difference is that $\mathfrak{m}$ as the tangent space at the origin of $K/L$ is the module of a representation of $\mathfrak{l}$ while after the modification $\mathfrak{m}$ is the module of a representation of $\mathfrak{l}\oplus \mathfrak{e}$.
Thus all the local homogeneous geometry of the modification $K\times E/L\times E$ is that  of $K/L$ which in addition is invariant under the representation
of $\mathfrak{e}$ on $\mathfrak{m}$.

\section{$N>16 ~ AdS_3\times_w M^8$ solutions in 11 dimensions}

\subsection{Fields}

We consider  warped AdS$_3$ backgrounds with internal space $M^8$, $\text{AdS}_3\times_w M^8$, with the most general allowed fluxes invariant under the symmetries of the AdS$_3$ subspace.
 The bosonic fields of 11-dimensional supergravity are   a metric $ds^2$ and a 4-form field strength $F$. Following the description of $\text{AdS}_3\times_w M^8$ backgrounds presented in \cite{mads},  these can be written as
\begin{align}
ds^2 &= 2 du(dr+rh)+ A^2 dz^2 + ds^2(M^8)~, \notag\\
F&= du \wedge (dr+rh) \wedge dz \wedge Q + X~,
\end{align}
where $(u,r,z)$ are the coordinates of AdS$_3$,
\begin{align}
h= -\frac{2}{\ell}\, dz - 2 A^{-1} dA~,
\end{align}
$\ell$ is the AdS$_3$ radius, $A$ is the warp factor which is a function of  $M^8$, and $Q$ and $X$ are a 1-form and 4-form on $M^8$, respectively.  The dependence of the fields on the AdS$_3$ coordinates $(u,r,z)$
is explicit while $ds^2(M^8), A, Q, X$ depend only on the coordinates $y^I$  of $M^8$.  Next we define a null-orthonormal frame as
\begin{align}\label{nullframe}
\bbe^+ = du~,~~ \bbe^-=dr + rh~,~~ \bbe^z = A dz~, ~~ \bbe^i = \bbe^i_{I} dy^I~,
\end{align}
with $ds^2(M^8) = \delta_{ij} \bbe^i \bbe^j$. The Bianchi identity $dF=0$ of  $F$ implies that
\begin{align}\label{mbianchi}
d(A^2Q)=0~,\quad dX=0~.
\end{align}
The field equations for $F$ give that
\bea
d*_{{}_8}X=-3 d\log A\wedge *_{{}_8}X-A^{-1} Q\wedge X~,
\label{coX}
\eea
and
\bea
d(A^{-1}*_{{}_8}Q)=-{1\over2} X\wedge X~,
\label{fqxx}
\eea
where  our Hodge duality conventions can be found in appendix A.
Similarly, the Einstein equation along $AdS_3$ gives rise to a field equation for the warp factor $A$
\begin{align}\label{warp}
A^{-1} \nabla^k \nabla_k A + 2 A^{-2} \nabla^k A \nabla_k A + \frac{2}{\ell^2 A^2} = \frac{1}{3A^2} Q^2 + \frac{1}{144} X^2~,
\end{align}
and the Einstein equation along $M^8$ reads
\begin{align}
R_{ij}^{(8)} = 3 A^{-1} \nabla_i\nabla_j A - \frac{1}{2} A^{-2} Q_i Q_j + \frac{1}{12} X^2_{ij} + \delta_{ij} \left(\frac{1}{6} A^{-2} Q^2 - \frac{1}{144} X^2 \right)~,
\end{align}
where $R^{(8)}_{ij}$ is the Ricci tensor of the internal manifold $M^8$. Note in particular that \eqref{warp} implies that $A$ is nowhere vanishing, provided that $A$ and all other fields are smooth.

\subsection{The Killing spinors}

Here we summarize the solution of the gravitino KSE of 11-dimensional supergravity   of \cite{mads} for warped $\text{AdS}_3\times_w M^8$  backgrounds. In this approach, the  KSE of 11-dimensional supergravity is first solved along the AdS$_3$ subspace and then the remaining independent KSEs
along the internal space $M^8$ are identified. The Killing spinors  can be expressed\footnote{The gamma matrices are always taken with respect to the null-orthonormal frame (\ref{nullframe}).}  as
\begin{align}\label{ks}
\epsilon  = \,&\sigma_+ +e^{-\frac{z}{\ell}} \tau_+ + \sigma_- + e^{\frac{z}{\ell}}  \tau_-  - \ell^{-1} u A^{-1} \Gamma_{+z} \sigma_- - \ell^{-1} r A^{-1} e^{-\frac{z}{\ell}} \Gamma_{-z} ~ \tau_+~,
\end{align}
where the dependence on the AdS$_3$ coordinates is explicit and $\sigma_\pm$ and $\tau_\pm$ are Majorana $\text{Spin}(10,1)$ spinors that depend only on the coordinates of
 $M^8$ and satisfy  the light-cone projections
\begin{align}
\Gamma_\pm \sigma_\pm = 0~, \quad \Gamma_\pm \tau_\pm = 0~.
\end{align}
The remaining independent KSEs on $M^8$ are
\begin{align}\label{graviKSE}
\nabla^{(\pm)}_i \sigma_\pm = 0~, \quad \nabla^{(\pm)}_i \tau_\pm = 0~,
\end{align}
and
\begin{align}\label{algKSE}
\Xi^{(\pm)}\sigma_{\pm}=0~, \quad (\Xi^{(\pm)} \pm \frac{1}{\ell})\tau_{\pm}=0~,
\end{align}
where
\begin{align}
\nabla_i^{(\pm)} &= \nabla_i \pm \frac{1}{2} \partial_i \log A - \frac{1}{288} \sgX_i + \frac{1}{36} \sX_i \mp \frac{1}{12} A^{-1} \Gamma_z \sgQ_i \pm \frac{1}{6} A^{-1} \Gamma_z Q_i~, \\
\Xi^{(\pm)} &= \mp \frac{1}{2\ell} - \frac{1}{2} \Gamma_z \slashed{\partial} A + \frac{1}{288} A \Gamma_z \sX \pm \frac{1}{6} \sQ~.
\end{align}
The conditions  \eqref{graviKSE} can be thought of as   the restriction of the gravitino KSE of 11-dimensional supergravity on  $M^8$ while  \eqref{algKSE} arises from the  integration of  the
gravitino KSE along the $AdS_3$ subspace.

To make a connection with the terminology used to describe the Killing superalgebras of AdS$_3$ backgrounds in section \ref{superxx},  the Killing spinors $\epsilon$ that depend only on the $\sigma_\pm$ type of spinors are in the left sector while those that depend on $\tau_\pm$ spinors are in the right sector.
The existence of  unrelated\footnote{In AdS$_n$, $n>3$, backgrounds the $\sigma_\pm$ and $\tau_\pm$ spinors are related by Clifford algebra operations.}
 $\sigma_\pm$ and $\tau_\pm$ types of spinors  is the reason that the Killing superalgebra $\mathfrak{g}$  of AdS$_3$ decomposes as $\mathfrak{g}=\mathfrak{g}_L\oplus \mathfrak{g}_R$.
Furthermore, it has been noted in \cite{mads} that if $\sigma_+$ and $\tau_+$ solve the KSEs (\ref{graviKSE}) and (\ref{algKSE}), so do
\begin{align}
\sigma_- = A \Gamma_{-z} \sigma_+~,\quad \tau_{-} = A \Gamma_{-z} \tau_+~.
\end{align}
Therefore the number of Killing spinors $N=N_L+N_R$ of AdS$_3$ backgrounds is always even, where $N_L$ and $N_R$  is the number of  Killing spinors of the left and right sector, respectively.

\subsection{For $N>16$ $AdS_3$ solutions $M^8$ is homogeneous}

\subsubsection{Factorization of Killing vectors}

It has been shown in \cite{superalgebra} that for  compact without boundary internal spaces $M^8$,  the even part of the Killing superalgebra $\mathfrak{g}_0$  decomposes into the algebra of symmetries of AdS$_3$ and those of the internal space $M^8$. This together with the homogeneity theorem of \cite{homogen} can be used to show that the internal space  $M^8$ is homogeneous for
$N>16$ backgrounds.

For AdS$_3$ backgrounds,  the condition \cite{superalgebra} for $\mathfrak{g}_0=\mathfrak{iso}(\text{AdS}_3)\oplus \mathfrak{t}_0$  is
\begin{align}\label{gamiz}
\langle \tau_+, \Gamma_{iz} \sigma_+ \rangle =0~,
\end{align}
for all $\sigma_+$ and $\tau_+$   spinors that satisfy (\ref{graviKSE}) and (\ref{algKSE}).  This can be derived using the compactness of $M^8$ as follows.
Setting $\Lambda= \sigma_+ + \tau_+$ and making use of the gravitino KSE \eqref{graviKSE}, one finds
\begin{align}
\nabla_i  \parallel \Lambda \parallel^2 = - \parallel \Lambda \parallel^2 A^{-1} \nabla_i A + \frac{1}{144} \langle \Lambda , \sgX_i \Lambda \rangle - \frac{1}{3} A^{-1} Q_i \langle \Lambda, \Gamma_z \Lambda \rangle~.
\label{middleqn}
\end{align}
Now, note that the algebraic KSE \eqref{algKSE} implies

\begin{align}
\frac{1}{\ell} (\sigma_+ -\tau_+) = (- \Gamma_{z}\slashed{d}A +\frac{A}{144} \Gamma_z \sX +\frac{1}{3} \sQ)~\Lambda~,
\end{align}
which, after multiplying by $A^{-1}\Gamma_{iz}$ and substituting back into (\ref{middleqn}), gives

\begin{align}\label{hopf1}
\nabla_i \parallel \Lambda \parallel^2=2 \ell^{-1} A^{-1}  \langle \tau_+, \Gamma_{iz} \sigma_+ \rangle~.
\end{align}
Furthermore, the gravitino KSE \eqref{graviKSE} also yields

\begin{align}
\nabla^i \left( A \langle \tau_+, \Gamma_{iz} \sigma_+ \rangle \right) = 0~.
\end{align}
Combining this with \eqref{hopf1}, one ends up with

\begin{align}
\nabla^2 \parallel \Lambda \parallel^2 + 2 A^{-1} \nabla^i A \nabla_i \parallel \Lambda \parallel^2 = 0~.
\end{align}
The Hopf maximum principle then implies that $\parallel \Lambda \parallel^2$ is constant, thus \eqref{hopf1} yields (\ref{gamiz}).

One consequence of (\ref{gamiz})  is that the linearly independent spinors  $\sigma_+$ and $\tau_+$,  on account of \eqref{algKSE},  are also orthogonal
\begin{align}
\langle \tau_+,  \sigma_+ \rangle =0~.
\end{align}
One can see this by taking $\langle \tau_+, \Xi^{(+)} \sigma_+ \rangle -\langle \sigma_+, (\Xi^{(+)} + \ell^{-1}) \tau_+ \rangle = 0$ and using \eqref{gamiz}.

\begin{table}
	\caption{8-dimensional compact, simply connected, homogeneous spaces}
	\centering
	\begin{tabular}{c l}
		\hline
		& $M^8=G/H$  \\  
		\hline
		(1)& $SU(3)$, group manifold\\
		(2)&$\frac{Sp(3)}{Sp(2)\times Sp(1)}=\mathbb{HP}^2$, symmetric space\\
		(3)& $\frac{SU(5)}{S(U(4)\times U(1))}=\mathbb{CP}^4$, symmetric space, not spin \\
		(4) & $\frac{Spin(9)}{Spin(8)}=S^8$, symmetric space\\
		(5) & $\frac{Sp(2)}{T^2}$, $T^2 \subset Sp(2)$ maximal torus\\
		(6) & $\frac{G_2}{SO(4)}$, symmetric space\\
		(7) & $\frac{SU(4)}{S(U(2)\times U(2))} = G_2(\mathbb{C}^4) =\frac{SO(6)}{SO(4)\times SO(2)} = G_2(\mathbb{R}^6)$, Grassmannian, symmetric space\\
		(8) & $\frac{SU(2)\times SU(2) \times SU(2)}{\Delta_{k,l,m}(U(1))}$\\
		(9) & $S^2\times S^6$ \\
		(10) & $S^2 \times \mathbb{CP}^3$ \\
		(11) & $S^2 \times \frac{SU(3)}{T^2}$\\
		(12) & $S^2 \times G_2(\mathbb{R}^5)$, not spin \\
		(13) & $S^3 \times S^5$ \\
		(14) & $S^3\times \frac{SU(3)}{SO(3)}$, not spin \\
		(15) & $S^4 \times S^4$ \\
		(16) & $S^4 \times \mathbb{CP}^2$, not spin \\
		(17) & $\mathbb{CP}^2 \times \mathbb{CP}^2$, not spin\\
		(18) & $S^2 \times S^2 \times S^4$ \\
		(19) & $ S^2 \times S^3 \times S^3$ \\
		(20) & $S^2\times S^2 \times S^2 \times S^2$ \\
		(21) & $S^2 \times S^2 \times \mathbb{CP}^2$, not spin \\[1ex]
		\hline
	\end{tabular}
	\label{table:8dim}
\end{table}

\subsubsection{$A$ is constant and  $M^8$ is homogeneous}

Let us define the spinor bilinear
\begin{align}
W_i = A\, \text{Im} \langle \chi_1 , \Gamma_{iz} \chi_2 \rangle~,
\end{align}
where $\chi$ either stands for $\sigma_+$ or $\tau_+$~. The gravitino KSE \eqref{graviKSE} then implies
\begin{align}
\nabla_{(i} W_{j)} =0~,
\end{align}
ie  $W$ is   a Killing vector\footnote{If the bilinear in (\ref{gamiz}) does not vanish, then the associated $W$ is not a Killing vector over the whole spacetime.} on $M^8$. From (\ref{gamiz}) it follows that the only non-vanishing Killing vector fields $W$  are those
that are constructed as bilinears of either $\sigma_+$ or  $\tau_+$ spinors.

As a consequence of the algebraic KSEs  \eqref{algKSE}, one has  $\text{Im} \langle \sigma_+^1, \Xi^{(+)} \sigma_+^2 \rangle =0$ and $\text{Im} \langle \tau_+^1, (\Xi^{(+)} +\ell^{-1}) \tau_+^2\rangle =0 $.  Expanding these,  one finds that
\begin{align}
i_W dA =0~,
\label{iwda}
\end{align}
where $W$ is a bilinear of either $\sigma_+$ or $\tau_+$ spinors.

As it has been mentioned,  (\ref{gamiz}) implies that the only non-vanishing Killing vectors $W$  on $M^8$ are  those constructed from either $\sigma_+$ or $\tau_+$ spinors.
Therefore (\ref{iwda}) will be  valid for all non-vanishing Killing vectors $W$ on $M^8$.
Suppose now that $N>16$. A similar argument to that used for the proof of the  homogeneity theorem in \cite{homogen} implies that the set of all Killing vectors  $W$ span the tangent space of $M^8$. Thus $A$ is constant and $M^8$ is homogeneous.  It should be noted that if (\ref{gamiz}) is not valid, then the vector fields $W$ in (\ref{iwda}) may not span all the Killing vectors on $M^8$.

Therefore we conclude that  all $N>16$ supersymmetric AdS$_3$  backgrounds are products AdS$_3\times M^8$, where $M^8$ is a homogeneous space.  In the analysis that follows, which includes that of AdS$_3$ backgrounds in type II 10-dimensional supergravities, we shall focus
only on such product spaces.

\subsection{Electric solutions do not preserve $16<N<32$ supersymmetries}

A consequence of the constancy of the warp factor is that it rules out the existence of electric solutions that preserve $16<N<32$ supersymmetries.  Indeed for electric
solutions $X=0$.  The algebraic KSE (\ref{algKSE}) on $\sigma_+$ reduces to
\bea
{1\over 3} \slashed{Q} \sigma_+={1\over\ell} \sigma_+~,
\eea
which implies the integrability condition
\bea
{1\over9} Q^2={1\over\ell^2}~.
\eea
On the other hand the field equation for the warp factor (\ref{warp})  yields ${1\over6} Q^2={1\over\ell^2}$ which is a contradiction as the radius of AdS$_3$ does not vanish, $\ell\not=0$.

\subsection{$N>16$ solutions with left only supersymmetry}

Suppose first that the solutions only have  left-hand supersymmetry.  In such a case, the Lie algebras that must act transitively  and effectively on the
internal spaces are
\bea
&&\mathfrak{so}(n)_L~,~~~n=9,\cdots, 15~, ~~~(N=2n)~;~~~
\cr
&&\mathfrak{u}(n)_L~,~~~n=5,6,7~,~~~(N=4n)~;~~~
\cr
&&(\mathfrak{sp}(3)\oplus \mathfrak{sp}(1))_L~,~~~N=24~,
\label{nr0}
\eea
where $N<32$ as there are no AdS$_3$ solutions which preserve maximal supersymmetry.  Furthermore solutions that preserve $N=30$ supersymmetries have already been excluded in \cite{11n30}.  An inspection of the list of homogeneous spaces reveals that the only possibility
that can occur is $S^8=\mathrm{Spin}(9)/\mathrm{Spin}(8)$ which can preserve 18 supersymmetries.  However $S^8$ is a symmetric space and there are no invariant 1- and 4-forms.
Thus $Q=X=0$ which in turn implies $F=0$.  This leads to a contradiction as the field equation for the warp factor cannot be satisfied.

\subsection{$N>16$ solutions with $N_R=2$}

For $N_R=2$  there are no right isometries and so all the symmetries of the internal space are  generated by $(\mathfrak{t}_L)_0$. The Lie algebras $(\mathfrak{t}_L)_0$ that  act transitively  and effectively on the
internal spaces are
\bea
&&\mathfrak{so}(n)_L~,~~~n=8,\cdots, 14~,~~~(N=2n+2)~;~~~
\cr
&&\mathfrak{u}(n)_L~,~~~n=4, \cdots,7~,~~~(N=4n+2)~;~~~
\cr
&&(\mathfrak{sp}(n)\oplus \mathfrak{sp}(1))_L~,~~~n=2,3~,~~~(N=8n+2)~;~~~
\cr
&&
\mathfrak{spin}(7)_L~~(N=18)~,
\label{nr2}
\eea
where the last case is associated with the Killing superalgebra  $\mathfrak{f}(4)$.
An inspection of the 8-dimensional homogeneous spaces in table \ref{table:8dim} reveals that there are only two possibilities that can occur
\bea
&&S^8=\mathrm{Spin}(9)/\mathrm{Spin}(8)~~(N=20)~,~~~~
\cr
&&
\mathbb{CP}^3\times S^2=Sp(2)/(Sp(1)\times U(1))\times Sp(1)/U(1)~~~(N=18)~.
\eea
Observe that $G_2(\mathbb{C}^4) =SU(4)/S(U(2)\times U(2))$ could have been included as a potential internal space of an AdS$_3$ background with $N=18$ supersymmetries provided that
it  admitted an effective $\mathfrak{u}(4)$ action.  However this is not the case as the rank of the isotropy group $S(U(2)\times U(2))$ is the same as that of $SU(4)$ and so it cannot admit
a modification such that $U(4)$ acts almost effectively on $G_2(\mathbb{C}^4)$.
For confirmation, we have also excluded this case with an explicit calculation which we shall not present here.

In addition  $\text{AdS}_3\times S^8$ can also be excluded as a solution  with an identical argument to the one we produced in the previous case with no right-handed supersymmetries. The remaining
case is investigated below.

  \subsubsection{$ \mathbb{CP}^3\times S^2=Sp(2)/(Sp(1)\times U(1))\times SU(2)/U(1)$}

For the analysis that follows, we use  the description of the geometry of the homogeneous space $Sp(2)/(Sp(1)\times U(1))$ presented in  \cite{ads4Ngr16}, where more details can be found.
  The metric on the internal space $ \mathbb{CP}^3\times S^2$ is
\bea
ds^2(M^8)= ds^2(\mathbb{CP}^3)+ ds^2 (S^2)~,
\eea
where
\bea
 ds^2(\mathbb{CP}^3)= a\, \delta_{ij}\bbl^i\bbl^j+ b\, \delta_{{\underline r}{\underline s}} \bbl^{\underline r}\bbl^{\underline s}~,~~~ds^2(S^2)=c\big( (\bbl^7)^2+(\bbl^8)^2\big)~,
 \eea
 and $(\ell^i, \ell^{\underline r})$, $i=1,\dots, 4$, ${\underline r}=1,2$ is a  left-invariant frame\footnote{In \cite{ads4Ngr16}, the left-invariant frame on $\mathbb{CP}^3$
 has been denoted as $(\ell^a, \ell^{\underline r})$ $a=1,\dots, 4$, ${\underline r}=1,2$ instead.}
   on $\mathbb{CP}^3$ and $(\ell^7, \ell^8)$ is a left-invariant frame
 on $S^2$. Moreover $a,b,c>0$ are constants.
As there are no invariant 1-forms $Q=0$.  The most general invariant 4-form is
\bea
X&=\frac{1}{2} \, \alpha_1 \, I^{(+)}_{3} \wedge I^{(+)}_{3}+ \alpha_2 \, \tilde{\omega}\wedge I^{(+)}_{3}+\alpha_3\, \sigma\wedge \tilde{\omega} + \alpha_4\, \sigma\wedge I^{(+)}_{3}  ~,
\eea
where $\alpha_1, \dots, \alpha_4$ are constants, $I^{(+)}_{3}=\ell^{12}+\ell^{34}$ and $\tilde{\omega}=\ell^{\underline {12}}$ are invariant 2-forms on $\mathbb{CP}^3$ whose properties can be found in \cite{ads4Ngr16} and $\sigma=\ell^{78}$.

\begin{table}[h]
\begin{center}
\vskip 0.3cm
 \caption{Decomposition of (\ref{433ccv})  into eigenspaces}
 \vskip 0.3cm

	\begin{tabular}{|c|c|}
		\hline
		$|J_1,J_2, J_3\rangle$&  relations for the fluxes\\
		\hline
		$|\pm,\pm,+\rangle$ & ${1\over 6}  \Big( {\alpha_1\over a^2} \mp  {2\alpha_2\over ba}-{\alpha_3\over bc} \pm {2\alpha_4\over ac}\Big) ={1\over \ell A} $ \\
\hline
		$|+,-, \pm\rangle$, $|-,+, \pm\rangle$& ${1\over 6}  \Big( {\alpha_1\over a^2}\pm{\alpha_3\over bc} \Big) =\pm {1\over \ell A}$  \\
		\hline
		$|\pm,\pm, -\rangle$ & ${1\over 6}  \Big( {\alpha_1\over a^2} \mp  {2\alpha_2\over ba}+{\alpha_3\over bc} \mp {2\alpha_4\over ac}\Big) =-{1\over \ell A} $ \\
		\hline
		
	\end{tabular}
\vskip 0.2cm
  \label{table43xcv}
 \end{center}
\end{table}

The closure and co-closure of $X$ give a relation between $\alpha_1$ and $\alpha_2$, and between $\alpha_3$ and $\alpha_4$, but they are not essential here.  Also $X\wedge X=0$ implies
that $\alpha_1 \alpha_3=0$.

On the other hand the algebraic KSE (\ref{algKSE}) can be written as
\bea
{1\over6} \Big({\alpha_1\over a^2} J_1 J_2-{\alpha_2\over ba} (J_1+J_2)- {\alpha_3\over bc} J_3+ {\alpha_4\over ac} (J_1+J_2) J_3\Big) J_1 J_2 J_3\sigma_+={1\over \ell A} \sigma_+~,
\label{433ccv}
\eea
where $J_1=\Gamma^{12{\underline {12}}}$, $J_2= \Gamma^{34{\underline {12}}}$ and $J_3=\Gamma^{78{\underline {12}}}$.  We have chosen the orientation such that
$\Gamma_z \sigma_+=-J_1J_2J_3\sigma_+$.  The decomposition of the algebraic KSE into the eigenspaces of $J_1, J_2$ and $J_3$ as well as the relations implied amongst the fluxes for each eigenspace can be found in table \ref{table43xcv}.

As each common eigenspace  of $J_1$, $J_2$ and $J_3$ has dimension two for solutions with $N>16$ supersymmetries one has always to consider either one of the eigenspinors  $|+,-, \pm\rangle$
and  $|-,+, \pm\rangle$ or all the eigenspinors $|\pm,\pm,+\rangle$ and $|\pm,\pm, -\rangle$. In the former  case, we have  that
\bea
{1\over36}\Big( {\alpha_1\over a^2}+{\alpha_3\over bc} \Big)^2={1\over \ell^2 A^2}~,
\eea
where we have chosen without loss of generality the eigenvalue $+1$  of $J_3$.
Taking the difference of the equation above with the warp factor field equation
\bea
{1\over12} \Big({\alpha_1^2\over a^4}+{2 \alpha_2^2\over b^2 a^2}+{\alpha_3^2\over b^2 c^2}+{2\alpha^2_4\over a^2 c^2}\Big)={1\over \ell^2 A^2}~,
\eea
we find that $\alpha_1=\alpha_2=\alpha_3=\alpha_4=0$, and so $X=0$,  which is a contradiction.  In the latter case we have that $\alpha_1=\alpha_2=\alpha_4=0$ and
${1\over6} {\alpha_3\over bc}=-{1\over \ell A}$. Comparing this with the warp factor field equation above again leads to a contradiction. There are no solutions with
internal space $Sp(2)/(Sp(1)\times U(1))\times SU(2)/U(1)$ that preserve $N>16$ supersymmetries.

\subsection{$N>16$ solutions with $N_R=4$}

The only right superalgebra that gives rise to 4 supersymmetries is $\mathfrak{osp}(2|2)$ which in turn leads to an $\mathfrak{so}(2)_R$ right-handed symmetry.
Therefore the Lie algebras that act both transitively and effectively on the internal spaces $M^8$  are
\bea
&&
\mathfrak{so}(n)_L\oplus \mathfrak{so}(2)_R~,~~~n=7,\cdots, 13~,~~~(N=2n+4)~;~~~~
\cr
&&
\mathfrak{u}(n)_L\oplus \mathfrak{so}(2)_R~,~~~n=4, 5, 6~,~~~(N=4n+4)~;
\cr
&&(\mathfrak{sp}(n) \oplus \mathfrak{sp}(1))_L\oplus \mathfrak{so}(2)_R~,~~~n=2,3~,~~~(N=8n+4)~;~~~~
\cr
&&
\mathfrak{spin}(7)_L \oplus \mathfrak{so}(2)_R~,~~~(N=20)~;~~~~
\cr
&&(\mathfrak{g}_2)_L\oplus \mathfrak{so}(2)_R~,~~~(N=18)~.
\label{nr4}
\eea
Up to a finite cover, the allowed  homogeneous spaces are
\bea
&&\mathrm{Spin}(7)/G_2\times S^1~,~~~(N=18, 20)~;~~~\mathrm{Spin}(8)/\mathrm{Spin}(7)\times S^1~,~~~(N=20)~;~~~
\cr
&&S^7\times S^1=U(4)/U(3)\times S^1~,~~~(N=20)~;~~~
\cr
&& S^7\times S^1=(Sp(2)\times Sp(1))/(Sp(1)\times Sp(1))\times S^1~,~~~(N=20)~;
\cr
&&S^4\times S^3\times S^1=\mathrm{Spin}(5)/\mathrm{Spin}(4) \times SU(2)\times S^1~,~~~(N=20)~.
\eea
Observe that all the cases that arise, up to discrete identifications,  are products of 7-dimensional homogeneous spaces with $S^1$. This is because it is not possible
to modify 8-dimensional homogeneous spaces which admit  an effective and transitive action of the $(\mathfrak{t}_L)_0$ Lie algebras in (\ref{nr4})
to homogeneous spaces which admit an effective and transitive action of $\mathfrak{t}_0=(\mathfrak{t}_L)_0\oplus \mathfrak{so}(2)_R$.  This is due to the fact
that for all candidate homogeneous spaces that can occur the rank of the isotropy group is the same as the rank of  $(\mathfrak{t}_L)_0$.

However a modification
has been used to include the homogeneous space $S^7\times S^1=(Sp(2)\times Sp(1))/(Sp(1)\times Sp(1))\times S^1$.  This is because an AdS$_3$  solution with internal space $Sp(2)/Sp(1)\times S^1$,
is expected to preserve $N=N_L+N_R=10+4=14<16$   supersymmetries as $\mathfrak{sp}(2)=\mathfrak{so}(5)$ and so should be discarded,  while with internal space $(Sp(2)\times Sp(1))/(Sp(1)\times Sp(1))\times S^1$ is expected to preserve $N=20$ supersymmetries as it is associated to the $(\mathfrak{sp}(n) \oplus \mathfrak{sp}(1))_L\oplus \mathfrak{so}(2)_R$ subalgebra in (\ref{nr4}) and therefore has been included.  A modification has also been used to include   $S^7\times S^1=U(4)/U(3)\times S^1$  as $S^7\times S^1=SU(4)/SU(3)\times S^1$ should have been discarded.

The coset space $\mathrm{Spin}(8)/\mathrm{Spin}(7)\times S^1$ can  immediately be excluded as the 4-form field strength $F$ is electric and we  have shown there are no electric
solutions which preserve $16<N\leq 32$ supersymmetries.  It remains to investigate the rest of the  cases.

\subsubsection{$\mathrm{Spin}(7)/G_2\times S^1$}

The metric on the homogeneous space $\mathrm{Spin}(7)/G_2\times S^1$  can be chosen as
\bea
ds^2(M^8)= ds^2(\mathrm{Spin}(7)/G_2)+ ds^2(S^1)= a\,  \delta_{ij} \bbl^i \bbl^j+b\, (\bbl^8)^2= \delta_{ij} \bbe^i \bbe^j+  (\bbe^8)^2~,
\eea
where a description of $\mathrm{Spin}(7)/G_2$ can be found in  \cite{ads4Ngr16} whose conventions we follow, $a,b>0$ are constants and $\ell^8$ is an invariant frame on $S^1$, $d\ell^8=0$.

The most general invariant fluxes are
\bea
Q=\gamma \bbe^8~,~~~X=\alpha\, *_{{}_7}\varphi+ \beta\, \bbe^8\wedge \varphi
\eea
where $*{}_7\varphi$ and $\varphi$ are the fundamental $G_2$ forms and $\alpha, \beta, \gamma$ are constants.  Furthermore, the Bianchi identity $dX=0$ implies that $ \beta=0$.

\vskip 0.5cm
\begin{table}[h]
\begin{center}
\vskip 0.3cm
 \caption{Decomposition of (\ref{Spin7G2_alg_KSE}) KSE into eigenspaces }
 \vskip 0.3cm
\begin{tabular}{|c|c|}
		\hline
		$|P_1,P_2,P_3\rangle$&  relations for the fluxes\\
		\hline
		$|+,+,+\rangle$, $|+,+,-\rangle$, $|-,+,+\rangle$, $|+,-,-\rangle$& $(-\frac{1}{6} \alpha \, \Gamma_z + \frac{1}{3} A^{-1} \gamma \Gamma_8  ) |\cdot\rangle = \frac{1}{\ell A} | \cdot \rangle$ \\
		$|-,+,-\rangle$, $|-,-,+\rangle$, $|-,-,-\rangle$&   \\
		\hline
		$|+,-,+\rangle$ & $(\frac{7}{6} \alpha \, \Gamma_z + \frac{1}{3} A^{-1} \gamma \Gamma_8  ) |\cdot\rangle = \frac{1}{\ell A} | \cdot \rangle$ \\
		\hline
		\end{tabular}
 \vskip 0.2cm
  \label{table3xxx}
 \end{center}
\end{table}

It is straightforward to observe that the investigation of the number of supersymmetries preserved  by the algebraic KSE is exactly the same as that for the AdS$_4$ backgrounds
 with internal space $\mathrm{Spin}(7)/G_2$ in \cite{ads4Ngr16},
where instead of $\Gamma_x$ we have $\Gamma_8$.  In particular, the algebraic KSE can be written as
\begin{align}\label{Spin7G2_alg_KSE}
\left(\frac{1}{6} \alpha \left( P_1 -P_2 +P_3 - P_1 \, P_2 \, P_3 - P_2 \, P_3 + P_1 \, P_3 -  P_1 \, P_2 \right) \Gamma_z + \frac{1}{3} \gamma \, A^{-1} \Gamma_8 \right) \sigma_+ = \frac{1}{\ell A} \sigma_+~,
\end{align}
where $\{P_1, P_2, P_3\}= \{ \Gamma^{1245}, \Gamma^{1267}, \Gamma^{1346}\}$ are mutually commuting, hermitian Clifford algebra operators with eigenvalues $\pm 1$.
The solutions of the algebraic KSE on the eigenspaces of $\{P_1, P_2, P_3\}$ have been tabulated in table \ref{table3xxx}.

To preserve $N>16$ supersymmetries, it is required to consider the subspace in table \ref{table3xxx} with  7 eigenspinors. The integrability condition of the remaining  algebraic KSE gives
\bea
{1\over36} \alpha^2+{1\over 9} A^{-2} \gamma^2={1\over\ell^2 A^2}~,
\eea
while the warp factor field equation implies
\bea
 {7\over12} \alpha^2+{1\over 6} A^{-2} \gamma^2={1\over\ell^2 A^2}~.
\eea
Clearly, these are mutually inconsistent.  So there are no AdS$_3$  solutions that preserve $N>16$ supersymmetries with internal space $\mathrm{Spin}(7)/G_2\times S^1$.

\subsubsection {$S^7\times S^1=U(4)/U(3)\times S^1$}\label{s7u4u3}

Let us briefly summarize  the homogeneous geometry of $S^7=U(4)/U(3)$ which is useful for our investigation of other cases below as well.
There is a left-invariant frame $(\ell^r, \ell^7)$, $r=1,\cdots, 6$,  on  $U(4)/U(3)$ such that the invariant metric can be written
as
\bea
ds^2(U(4)/U(3))= a\, (\ell^7)^2+b\, \delta_{rs} \ell^r \ell^s~,
\label{metru4u3}
\eea
where $a,b>0$ are constants.  The invariant forms on  $U(4)/U(3)$ are generated by the invariant 1-form $\ell^7$ and the
2-form $\omega$ which can be chosen as
\bea
\omega=\ell^{12}+\ell^{34}+\ell^{56}~.
\label{omegau4u3}
\eea
Furthermore
\bea
d\ell^7=\omega~.
\label{diffu4u3}
\eea
For more details see eg \cite{ads4Ngr16}, where the homogeneous geometry of $SU(4)/SU(3)$ is also described.

Turning to the investigation at hand, the metric on $U(4)/U(3)\times S^1$ can be written as
\bea
ds^2(M^8)=ds^2(U(4)/U(3)) + ds^2(S^1)~,~~~~ds^2(S^1)=c \, (\ell^8)^2~,
\eea
where $ds^2(U(4)/U(3))$ is as in (\ref{metru4u3}), $\ell^8$ is the invariant frame on $S^1$, $d\ell^8=0$,  and $c>0$ is constant.

The most general invariant fluxes $Q$ and $X$ that satisfy the Bianchi identities (\ref{mbianchi}),  $dX=dQ=0$, are
\begin{align}
X =  \frac{1}{2}\, \alpha\, \omega\wedge\omega ~,~~~Q=\beta\,  \ell^8~,
\end{align}
where  $\alpha,\beta $ are constants.

Next consider the Einstein equation along $S^1$.  As $X$ does not have non-vanishing components along $S^1$ and the metric factorizes into that of  $U(4)/U(3)$ and $S^1$, we have
\begin{align}
R_{88}^{(8)} =  -  \frac{1}{3} A^{-2} Q^2 - \frac{1}{144} X^2 ~,
\label{r88s1}
\end{align}
where  $R_{88}^{(8)}$ is the Ricci tensor along $S^1$. This must vanish, $R_{88}^{(8)}=0$.  Thus  $Q=X=0$.  Then the warp factor field equation cannot be satisfied and so there are no AdS$_3$
solutions with internal space $U(4)/U(3)\times S^1$.

\subsubsection{$S^7\times S^1=(Sp(2)\times Sp(1))/(Sp(1)\times Sp(1))\times S^1$} \label{sp2sp1sp1sp1s1}

The modification of $Sp(2)/Sp(1)$ to $(Sp(2)\times Sp(1))/(Sp(1)\times Sp(1))$ has already been described in section \ref{modifx} and in particular in (\ref{modifsp2}).
The geometry of this homogeneous space is a special case of that of  $Sp(2)/Sp(1)$.  In particular, the invariant forms on $(Sp(2)\times Sp(1))/(Sp(1)\times Sp(1))$
are those on $Sp(2)/Sp(1)$ which are invariant under both $Sp(1)$'s in the isotropy group.

Using the notation in \cite{ads4Ngr16}, we introduce a left-invariant frame $(\ell^a, \ell^r )$ on $(Sp(2)\times Sp(1))/(Sp(1)\times Sp(1))$, where $a=1,2,3,4$ and $r=5,6,7$.  Then imposing  invariance under both $Sp(1)$'s, one finds that
there are no invariant 1- and 2-forms on $(Sp(2)\times Sp(1))/(Sp(1)\times Sp(1))$.  However there are two invariant 3-forms and  two invariant 4-forms given by
\bea
\sigma={1\over 3!} \epsilon_{rst} \ell^{rst}~,~~~\tau=\ell^r\wedge I_r^{(+)}~,
\eea
\bea
\rho=\delta^{rs} \rho_{rs}={1\over2}\delta^{rs} \epsilon_{rpq} \ell^{pq}\wedge I_s^{(+)}~,~~~\psi={1\over4!}\epsilon_{abcd} \ell^{abcd}~,
\eea
respectively, where $I_r^{(+)}={1\over2} (I_r^{(+)})_{ab}\ell^{ab}$ and $\big( (I_r^{(+)})_{ab}\big)$ is a basis of self-dual 2-forms on $\bR^4$.  Moreover
\bea
d\sigma={1\over2} \rho~,~~~d\tau=3 \psi- \rho~,~~~d\psi=d\rho=0~.~~~
\eea

After imposing the Bianchi identities $dQ=dX=0$, the  most general fluxes can be written as
\bea
X=\alpha_1 \psi+ \alpha_2 \rho~,~~~Q= \beta \ell^8~,
\eea
where $\ell^8$ is an invariant frame on $S^1$, $d\ell^8=0$.

The metric can be chosen as
\bea
ds^2 = f\, \delta_{ab} \ell^a \ell^b + h\, \delta_{rs} \ell^r \ell^s+ p\, (\ell^8)^2~,
\label{metrsp2sp1sp1sp1s1}
\eea
where $f,h,p>0$ are constants.  Substituting the metric and fluxes into the Einstein equation along the $S^1$ direction, we find again (\ref{r88s1}) which implies
 $Q=X=0$.  So there are no AdS$_3$ solutions with internal space $(Sp(2)\times Sp(1))/(Sp(1)\times Sp(1))\times S^1$.

\subsubsection{$S^4\times S^3\times S^1=\mathrm{Spin}(5)/\mathrm{Spin}(4) \times SU(2)\times S^1$}

The metric can be chosen as
\bea
ds^2(M^8)= a\, \delta_{ij} \ell^i \ell^j+ b_{rs} \ell^r \ell^s+ c\, (\ell^8)^2~,
\eea
where
\bea
ds^2(S^4)=a\, \delta_{ij} \ell^i \ell^j~,~~~ds^2(S^3)=b_{rs}\, \ell^r \ell^s~,~~~ds^2(S^1)=c\, (\ell^8)^2~,
\eea
and where $a, c>0$ are constants, $b=(b_{rs})$ is a constant symmetric positive definite matrix.  $(\ell^i)$, $i=1,2,3,4$, is a left-invariant frame on $S^4$ viewed
as a $\mathrm{Spin}(5)/\mathrm{Spin}(4)$ symmetric space and $(\ell^r)$, $r=5,6,7$, is a left-invariant frame on the group manifold $S^3$ with
\bea
d\ell^r={1\over2} \epsilon^r{}_{st} \ell^s\wedge \ell^t~,
\eea
and $\ell^8$ is an invariant  frame on $S^1$, $d\ell^8=0$.  Note that $\ell^r$ can be chosen up to an $SO(3)$ transformation.  This can be used to choose $b$ without loss of generality  to be
diagonal.

The most general invariant fluxes are
\bea
X&=&\alpha_1\, \ell^{1234}+ \alpha_2\, \ell^{5678}~,~~~
Q=\beta_1 \ell^8+ \gamma_r \ell^r~.
\eea
As the Bianchi identities require that $dQ=0$, one finds that $\gamma_r=0$.  Set $\beta=\beta_1$.  As $Q$ is also co-closed, we have that $X\wedge X=0$ which in turn gives
$\alpha_1\alpha_2=0$.

Suppose first that $\alpha_1=0$.  In that case, the algebraic KSE can be written as
\bea
\Big({1\over 6} {\alpha_2\over \sqrt {c b_1b_2b_3}} J_1+ {1\over3} {\beta \over A \sqrt{c}} J_2\Big)\sigma_+={1\over \ell A} \sigma_+~,
\eea
where $J_1=\Gamma^{5678} \Gamma_z$ and $J_2=\Gamma^8$ are commuting hermitian Clifford algebra operators and $b=\mathrm{diag} (b_1, b_2,b_3)$. To find solutions with $N>16$ supersymmetries we have to consider
at least two of the common eigenspaces of $J_1$ and $J_2$ each of which has dimension 4.  This is possible if either $\alpha_2$ or  $\beta$ vanishes.  If $\alpha_2=0$,  then $X=0$ and the solution
is purely electric.  Such solutions cannot preserve $N>16$ supersymmetries.  On the other hand if $\beta=0$, the integrability condition of the KSE implies that
\bea
{1\over 36} {\alpha^2_2\over c\, b_1 b_2b_3}={1\over \ell^2 A^2}~.
\eea
Comparing this with the warp factor field equation, it leads to an inconsistency.  Thus there are no such AdS$_3$ solutions which preserve $N>16$ supersymmetries with internal space $\mathrm{Spin}(5)/\mathrm{Spin}(4) \times SU(2)\times S^1$.

Suppose now that $\alpha_2=0$.  In such a case $X$ does not have components along $S^1$.  As a result the Einstein equations along $S^1$ can be written as in (\ref{r88s1})
and so $X=Q=0$.  There are no such AdS$_3$ solutions with internal space $\mathrm{Spin}(5)/\mathrm{Spin}(4) \times SU(2)\times S^1$.

\subsection{$N>16$ solutions with $N_R=6$}

The only right-handed superalgebra with 6 odd generators is $\mathfrak{osp}(3|2)$.  This   gives rise to an $\mathfrak{so}(3)_R$ action on the internal space.
Therefore the symmetry algebras that act transitively and effectively on  the internal spaces  are
\bea
&&\mathfrak{so}(n)_L\oplus \mathfrak{so}(3)_R~,~~~n=6, \cdots, 12~,~~~(N=2n+6)~;~~~~
\cr
&&\mathfrak{u}(n)_L\oplus \mathfrak{so}(3)_R~,~~~n=3, 4, 5, 6~,~~~(N=4n+6)~;
\cr
&&(\mathfrak{sp}(n) \oplus \mathfrak{sp}(1))_L\oplus \mathfrak{so}(3)_R~,~~~n=2,3~,~~~(N=8n+6)~;~
\cr
&&
\mathfrak{spin}(7)_L \oplus \mathfrak{so}(3)_R~,~~~(N=22)~;~~~~
\cr
&&(\mathfrak{g}_2)_L\oplus \mathfrak{so}(3)_R~,~~~(N=20)~.
\label{nr6}
\eea
An inspection of the homogeneous spaces in table \ref{table:8dim} reveals that up to a finite covering these are either $M^6\times S^2$ or $M^5\times S^3$, where $M^6$ and $M^5$ are
homogeneous 6- and 5-dimensional spaces.  So we have
\bea
&&S^6\times S^2=\mathrm{Spin}(7)/\mathrm{Spin}(6)\times SU(2)/U(1)~,~~~(N=20, 22)~;~~~
\cr
&&\mathbb{CP}^3\times S^2=SU(4)/S(U(1)\times U(3))\times SU(2)/U(1)~,~~~(N=18)~;
\cr
&&S^5\times S^3=\mathrm{Spin}(6)/\mathrm{Spin}(5)\times SU(2)~,~~~(N=18)~;~~~
\cr
&&S^5\times S^3=U(3)/U(2) \times SU(2)~,~~~(N=18)~;
\cr
&&S^4\times S^2\times S^2=\mathrm{Spin}(5)/\mathrm{Spin}(4)\times SU(2)/U(1)\times  SU(2)/U(1)~,~~~(N=22)~;
\cr
&&S^6\times S^2=G_2/SU(3)\times SU(2)/U(1)~,~~~(N=20)~.
\eea
The homogeneous space $SU(3)/T^2\times SU(2)/U(1)$ has been excluded as there is no modification that can be made such that  $U(3)$ can act almost  effectively on it.
Nevertheless we have performed the analysis to demonstrate that it cannot be the internal space of an AdS$_3$ solution that preserves $N>16$ supersymmetries.
 On the other hand $SU(4)/S(U(1)\times U(3))\times SU(2)/U(1)$ has been included because $\mathfrak{su}(4)=\mathfrak{so}(6)$
and so $\mathbb{CP}^3$ admits an $\mathfrak{so}(6)$ effective and transitive action giving rise to  $N=18$ supersymmetries with $N_L=12$ and $N_R=6$.  $SU(4)/S(U(1)\times U(3))\times SU(2)/U(1)$ could have been considered as a background that preserves $20$ supersymmetries as well but it cannot be modified to admit an effective
$\mathfrak{u}(4)$ action.

The homogeneous spaces $S^6\times S^2=\mathrm{Spin}(7)/\mathrm{Spin}(6)\times SU(2)/U(1)$ and
$S^5\times S^3=\mathrm{Spin}(6)/\mathrm{Spin}(5)\times SU(2)$ can immediately be excluded as giving potential solutions. For $S^6\times S^2$, $X=Q=0$ and so the warp factor field equation cannot be satisfied.  The same is the case for  $S^5\times S^3$ after applying the Bianchi identity $dQ=0$ to show that $Q=0$.

\subsubsection{$\mathbb{CP}^3\times S^2=SU(4)/S(U(1)\times U(3))\times SU(2)/U(1)$}

This homogeneous space is considered as an internal space because $\mathfrak{su}(4)=\mathfrak{so}(6)$ and so it may give rise to a solution which
preserves 18 supersymmetries.  The most general invariant  metric in the conventions of  \cite{ads4Ngr16} is
\begin{align}
ds^2(M^8)= ds^2(S^2)+ds^2(\mathbb{CP}^3)= a\, \delta_{ij} \ell^i \ell^j +b (\delta_{rs} \ell^r \ell^s + \delta_{\tilde{r}\tilde{s}} \ell^{\tilde{r}} \ell^{\tilde{s}})~,
\end{align}
where $(\ell^i)$, $i=7,8$, is a left-invariant frame on $S^2$ and $(\ell^r, \ell^{\tilde{r}})$  $r,\tilde{r}=1,2,3$, is a left-invariant frame on  $\mathbb{CP}^3$ and $a,b>0$ are constants. The invariant forms are generated by the volume form on $S^2$
\begin{align}
\sigma = \frac{1}{2} a\, \epsilon_{ij} \ell^i \wedge \ell^j~,
\end{align}
and the K\"ahler form on $\mathbb{CP}^3$
\begin{align}
\omega = b\, \delta_{r\tilde{s}} \ell^r \wedge \ell^{\tilde{s}}~.
\end{align}
Hence the most general invariant  fluxes are
\begin{align}
Q=0~, \quad X= \alpha\, \frac{1}{2} \omega\wedge \omega + \beta\, \sigma\wedge\omega~.
\end{align}
The Bianchi identities are trivially satisfied but the field equation for $Q$ gives  the condition
\begin{align}
X\wedge X = \alpha \beta~ \sigma\wedge\omega\wedge\omega\wedge\omega = 0~.
\end{align}
Therefore, either $\alpha=0$ or $\beta=0$. It remains to investigate the KSEs.

\underline{$\beta=0$}

For $\beta=0$, the flux $X$ is simply $X= \frac{1}{2} \alpha\, \omega\wedge\omega$. Going to an  orthonormal frame, in which the K\"ahler form is $\omega= \bbe^{12} + \bbe^{34} +\bbe^{56}$, we find for the algebraic KSE \eqref{algKSE}
\begin{align}\label{alphakse}
\frac{\alpha}{6} (J_1+J_2 - J_1 J_2) \Gamma_z \sigma_+ = \frac{1}{\ell A} \sigma_+~,
\end{align}
where $J_1=\Gamma^{1234}$ and $J_2= \Gamma^{1256}$ are mutually commuting Clifford algebra operators with eigenvalues $\pm1$. The decomposition in terms of the common eigenspaces is summarised in table \ref{tablealphaksex}. A similar analysis applies to $\tau_+$, except that the right-hand side is $-1/(\ell A)$.

\begin{table}[h]
	\begin{center}
		\vskip 0.3cm
		\caption{Decomposition of \eqref{alphakse} KSE into eigenspaces}
		\vskip 0.3cm
		\begin{tabular}{|c|c|c|}
			\hline
			&$|J_1,J_2\rangle$&  relations for the fluxes\\
			\hline
			(1)&$|+,+\rangle$, $|+,-\rangle$, $|-,+\rangle$ & $\frac{\alpha}{6} \Gamma_z |\cdot\rangle = \frac{1}{\ell A} | \cdot \rangle$ \\
\hline
			(2)& $|-,-\rangle$ &  $-\frac{\alpha}{2} \Gamma_z |\cdot\rangle = \frac{1}{\ell A} | \cdot \rangle$ \\
			\hline
		\end{tabular}
		\vskip 0.2cm
		\label{tablealphaksex}
	\end{center}
\end{table}

To find solutions that preserve  $N>16$ supersymmetries, one has to choose  spinors from the  eigenspaces (1) in table \ref{tablealphaksex}. In such a case,  the integrability condition of the remaining $\Gamma_z$ projection on the spinors  is
\begin{align}
\frac{\alpha^2}{36} = \frac{1}{\ell^2 A^2}~,
\end{align}
whereas the field equation for the warp factor \eqref{warp} requires
\begin{align}
\frac{\alpha^2}{4} = \frac{1}{\ell^2 A^2}~.
\end{align}
Thus there is a contradiction and there are no AdS$_3$ solutions preserving $N>16$ supersymmetries.

\underline{$\alpha=0$}

For $\alpha=0$, the 4-form flux becomes $X=\beta \sigma\wedge\omega$. Going to an orthonormal frame, in which $\omega = \bbe^{12} + \bbe^{34} +\bbe^{56}$ and $\sigma = \bbe^{78}$, we find for the algebraic KSE \eqref{algKSE}
\begin{align}\label{betakse}
-\frac{\beta}{6} (J_1 + J_2 + J_3) J_1J_2J_3 \sigma_+ = \frac{1}{\ell A} \sigma_+~,
\end{align}
where the Clifford algebra operators $J$ are defined as
\begin{align}
J_1 = \Gamma^{1278}~,~~~J_2= \Gamma^{3478}~,~~~J_3= \Gamma^{5678}~,
\end{align}
and
\bea
\Gamma_z=-J_1J_2J_3~.
\eea
The decomposition of the algebraic KSE \eqref{betakse} into the eigenpaces of these mutually commuting Clifford algebra operators is illustrated in table \ref{tablebetaksexx}. A similar analysis applies to the $\tau_+$ spinors with the right-hand side replaced by $-\frac{1}{\ell A}$.

\begin{table}[h]
	\begin{center}
		\vskip 0.3cm
		\caption{Decomposition of \eqref{betakse} KSE into eigenspaces}
		\vskip 0.3cm
		\begin{tabular}{|c|c|c|}
			\hline
			&$|J_1,J_2,J_3\rangle$&  relations for the fluxes\\
			\hline
			(1)&$|\pm,\pm,\mp\rangle$, $|\pm,\mp,\pm\rangle$, $|\mp,\pm,\pm\rangle$ & $\frac{\beta}{6}  = \frac{1}{\ell A} $ \\
\hline
			(2)& $|\pm,\pm,\pm\rangle$ &  $-\frac{\beta}{2}  = \frac{1}{\ell A} $ \\
			\hline
		\end{tabular}
		\vskip 0.2cm
		\label{tablebetaksexx}
	\end{center}
\end{table}

For solutions to preserve  $N>16$ supersymmetries, we need to consider the eigenspinors given in row   (1) of table \ref{tablebetaksexx}.  This gives
\begin{align}
\frac{\beta^2}{36} = \frac{1}{\ell^2 A^2}~,
\end{align}
while the field equation for the warp factor \eqref{warp} leads to
\begin{align}
\frac{\beta^2}{4} = \frac{1}{\ell^2 A^2}~.
\end{align}
Clearly this is a  contradiction.  There are no AdS$_3$ backgrounds that preserve $N>16$ supersymmetries with internal space $SU(4)/S(U(1)\times U(3))\times SU(2)/U(1)$.

\subsubsection{$S^5\times S^3=U(3)/U(2) \times SU(2)$} \label{u3u2su2}

The  geometry on $S^5$ as a $U(3)/U(2)$ homogeneous space can be described in a similar way as that for $S^7=U(4)/U(3)$ which can be found in section \ref{s7u4u3}. In particular the metric is
  \bea
ds^2(U(3)/U(2))= a\, (\ell^5)^2+b\, \delta_{rs} \ell^r \ell^s~,~~~r,s=1,2,3,4~,
\label{metru3u2}
\eea
where $a,b>0$ are constants.  The invariant forms on  $U(3)/U(2)$ are generated by the  1-form $\ell^5$ and the
2-form $\omega=\ell^{12}+\ell^{34}$.  Again  $d\ell^5=\omega$.

 The existence of AdS$_3$ solutions with internal space $S^5\times S^3$ can be ruled out with a cohomological argument.  Indeed let $\ell^i$ be a left-invariant frame on $S^3$ such that
  \bea
  d\ell^i={1\over2} \epsilon^i{}_{jk} \ell^j\wedge \ell^k~,~~~i,j,k=1,2,3~.
  \eea
  The most general invariant 1-form $Q$ can be written as
\bea
Q= \alpha \bbl^5+ \beta_r \bbl^r~.
\eea
 The Bianchi identity, $dQ=0$, in \eqref{mbianchi} implies that $\alpha=\beta_r=0$.
So, we have  $Q=0$.

Furthermore, the Bianchi identities \eqref{mbianchi} also imply that $dX=0$, and as $Q=0$ the field equation \eqref{coX} also implies that $d*X=0$.  Thus $X$ is harmonic and  represents a class in $H^4(S^5\times S^3)$.
However $H^4(S^5\times S^3)=0$ and so $X=0$.  This leads to a contradiction as the field equation for the warp factor \eqref{warp} cannot be satisfied.

Note that the above calculation rules out the existence of AdS$_3$ solutions with internal space $S^5\times S^3=\mathrm{Spin}(6)/\mathrm{Spin}(5) \times SU(2)$ as this is a special case of the
background examined above.

\subsubsection{$S^4\times S^2\times S^2=\mathrm{Spin}(5)/\mathrm{Spin}(4)\times SU(2)/U(1)\times  SU(2)/U(1)$}

The most general invariant metric is
\bea
ds^2(M^8)&=& ds^2(S^4)+ ds^2(S^2)+ ds^2(S^2)
\cr
&=& a\, \delta_{ij} \ell^i \ell^j+ b\, \left((\ell^5)^2+(\ell^6)^2\right)+ c\,\left(\ell^7)^2+(\ell^8)^2\right)~,
\eea
where $a,b,c>0$ are constants, $\ell^i$, $i=1,2,3,4$, is a left-invariant frame on $S^4$ viewed as the symmetric space $\mathrm{Spin}(5)/\mathrm{Spin}(4)$, and $(\ell^5, \ell^6)$ and $(\ell^7, \ell^8)$ are left-invariant frames on the two $S^2$'s, respectively.

As there are no invariant 1-forms $Q=0$.  Moreover $X$ can be written as
\bea
X=\alpha_1 \ell^{1234}+\alpha_2 \ell^{5678}~.
\eea
As $X\wedge X=0$, which follows from the field equation of $Q$, we have that $\alpha_1\alpha_2=0$.  If $\alpha_2=0$, then the integrability condition of the algebraic KSE will
give
\bea
{1\over 36} {\alpha_1^2\over a^4}={1\over \ell^2 A^2}~.
\eea
Comparing this with the field equation of the warp factor leads to a contradiction.  This is also the case if instead $\alpha_1=0$.  There are no supersymmetric AdS$_3$ solutions with internal space  $S^4\times S^2\times S^2$.

\subsubsection{$S^6\times S^2=G_2/SU(3)\times SU(2)/U(1)$}

The existence of AdS$_3$ solutions with  $G_2/SU(3)\times SU(2)/U(1)$ internal space can be ruled out by a cohomological argument.  Observe that $\mathfrak{su}(3)$ acts on $\mathfrak{m}$ with the $[\bf 3]_R={\bf 3}\oplus \bar{\bf{3}}$ representation. Using this, one concludes that there are no  invariant 1-forms on $M^8$ and so  $Q=0$.  In such a case $X$ is both closed and co-closed and so harmonic.   However,  $H^4(M^8)=0$ as $M^8=S^6\times S^2$ and so $X=0$.  This  in turn leads to a  contradiction as the field equation
for the warp factor cannot be satisfied.

\subsection{$N>16$ solutions with $N_R=8$}

The right-handed superalgebras with 8 supercharges are $\mathfrak{osp}(4|2)$, $\mathfrak{D}(2,1, \alpha)$ and $\mathfrak{sl}(2|2)/1_{4\times 4}$.
These give rise to right-handed isometries with Lie algebras $\mathfrak{so}(4)_R$, $(\mathfrak{so}(3)\oplus \mathfrak{so}(3))_R$ and $\mathfrak{su}(2)_R$,
respectively.  In the latter case there can also be up to three additional central generators. As $\mathfrak{so}(4)_R=(\mathfrak{so}(3)\oplus \mathfrak{so}(3))_R$, it suffices to consider  $(\mathfrak{so}(3)\oplus \mathfrak{so}(3))_R$ and $\mathfrak{su}(2)_R$, and in the latter case include up to 3 central generators.   Furthermore as $N>16$, one has
 $N_L> 8$. Collecting the above and using the results of table \ref{table:ads3ksa},  the allowed algebras that act transitively and effectively on the internal space are the following.
 \bea
&& \mathfrak{so}(n)_L\oplus (\mathfrak{t}_R)_0~,~~~n=5,\dots,11~,~~~(N=2n+8)~;
\cr
&&\mathfrak{u}(n)_L\oplus (\mathfrak{t}_R)_0~,~~~n=3,4,5~,~~~(N=4n+8)~;
\cr
&&(\mathfrak{sp}(2)\oplus\mathfrak{sp}(1))_L\oplus (\mathfrak{t}_R)_0~,~~~(N=24)~;
\cr
&&\mathfrak{spin}(7)_L\oplus (\mathfrak{t}_R)_0~,~~~(N=24)~;~~~(\mathfrak{g}_2)_L\oplus (\mathfrak{t}_R)_0~,~~~(N=22)~,
\label{nr8}
\eea
where $(\mathfrak{t}_R)_0$ is either $(\mathfrak{so}(3)\oplus \mathfrak{so}(3))_R$ or $\mathfrak{su}(2)_R\oplus \mathfrak{c}_R$ with  $\mathfrak{c}_R$ spanned by up to 3 central generators.
The homogeneous spaces that can admit a transitive and an effective action  by the above Lie algebras  have been tabulated in table \ref{nr8table}.

\begin{table}[h]
	\begin{center}
		\vskip 0.3cm
		\caption{Homogeneous spaces for $N_R=8$}
		\vskip 0.3cm
\scalebox{0.8}{%
		\begin{tabular}{|c|c|c|}
			\hline
$\mathfrak{t}_0$& Homogeneous spaces& $N$\\
\hline
			$\mathfrak{so}(5)_L\oplus (\mathfrak{t}_R)_0$&  $ S^4\times S^2\times S^2=\mathrm{Spin}(5)/\mathrm{Spin}(4)\times SU(2)/U(1)\times  SU(2)/U(1)$& 18\\
			& $S^4\times S^2\times
 T^2=\mathrm{Spin}(5)/\mathrm{Spin}(4)\times SU(2)/U(1)\times  T^2 $ & 18\\
			  &  $
 \mathbb{CP}^3\times S^2=Sp(2)/(Sp(1)\times U(1))\times SU(2)/U(1) $&18 \\
  &$S^7\times S^1=(Sp(2)\times Sp(1))/(Sp(1)\times Sp(1))\times S^1$&18\\
			\hline
$\mathfrak{so}(6)_L\oplus (\mathfrak{t}_R)_0$&$S^5\times S^3=\mathrm{Spin}(6)/\mathrm{Spin}(5)\times SU(2)$&20\\
&$
 S^5\times S^2\times
 S^1=\mathrm{Spin}(6)/\mathrm{Spin}(5)\times SU(2)/U(1)\times  S^1$&20\\
 &$\mathbb{CP}^3\times S^2=SU(4)/S(U(3)\times U(1))\times SU(2)/U(1)$ &20\\
 \hline
 $\mathfrak{so}(7)_L\oplus (\mathfrak{t}_R)_0$&$S^6\times S^2=\mathrm{Spin}(7)/\mathrm{Spin}(6)\times SU(2)/U(1)$&22, 24\\
 \hline
 $\mathfrak{u}(3)_L\oplus (\mathfrak{t}_R)_0$&$SU(3)^{k,l}=(SU(3)\times SU(2)\times U(1)/(SU(2)\times \Delta_{k,l}U(1))$&20\\
 &$S^5\times S^3=U(3)/U(2)\times SU(2)$&20\\
 &$S^5\times S^2\times S^1=U(3)/U(2)\times SU(2)/U(1)\times S^1$&20\\
 &$N^{k,l,m}\times S^1=\frac{U(1)\times SU(2) \times SU(3) }{\Delta_{k,l,m}((U(1)^2)\cdot (1\times SU(2))}\times S^1$&20\\
 \hline
 $\mathfrak{sp}(2)\oplus \mathfrak{sp}(1))_L\oplus (\mathfrak{t}_R)_0$&$S^4\times S^2\times S^2=\mathrm{Spin}(5)/\mathrm{Spin}(4)\times SU(2)/U(1)\times SU(2)/U(1)$&24\\
 &$S^4\times S^3\times S^1=\mathrm{Spin}(5)/\mathrm{Spin}(4)\times (SU(2)\times SU(2))/SU(2)\times S^1$&24\\

 \hline
 $( \mathfrak{g}_2)_L\oplus (\mathfrak{t}_R)_0$&$S^6\times S^2=G_2/SU(3)\times SU(2)/U(1)$&22\\
 \hline
 \end{tabular}}
		\vskip 0.2cm
		\label{nr8table}
	\end{center}
\end{table}

A detailed examination of the homogeneous spaces that may give rise to supersymmetric AdS$_3$ solutions with $N_R=8$ reveals that the only cases that have not
been investigated so far are $S^4\times S^2\times T^2$, $S^5\times S^2\times S^1$ with $S^5$ either $\mathrm{Spin}(6)/\mathrm{Spin}(5)$ or $U(3)/U(2)$, $SU(3)$ and
$N^{k,l,m}\times S^1$.  The remaining homogeneous spaces have already been excluded as internal spaces  in the analysis of AdS$_3$ backgrounds with  $N_R<8$ backgrounds.    The presence of additional right-handed supersymmetries here for $N_R=8$
are not sufficient to bring these backgrounds into the range  of $N>16$ supersymmetries.  So again they are excluded as solutions.

 \subsubsection{$ S^4\times S^2\times
 T^2=\mathrm{Spin}(5)/\mathrm{Spin}(4)\times SU(2)/U(1)\times  T^2$}

  The most general invariant metric is
\begin{align}
ds^2 (M^8)=ds^2(S^4)+ ds^2(S^2)+ds^2(T^2)=a~ \delta_{rs} \ell^r \ell^s+ b~ \delta_{\hat a\hat b} \ell^{\hat a} \ell^{\hat b} + c_{\tilde{a}\tilde{b}} \ell^{\tilde{a}} \ell^{\tilde{b}} ~,
\end{align}
where $\ell^r$, $r=1,...,4$, is a left-invariant frame on $S^4$, $ \ell^{\hat a}$,  $\hat a = 5,6$ is a left-invariant frame on $S^2$ and  $\ell^{\tilde{a}}$,  $\tilde{a}=7,8$,  is a left
 invariant frame on $T^2$, $d\ell^{\tilde{a}}=0$, and $a,b>0$ are constants and $(c_{\tilde a \tilde b})$ is a positive definite matrix. The invariant forms on this $M^8$ are generated by
   $\ell^{\tilde{a}}$ and the top forms on $S^4$ and $S^2$.  Hence the 4-form flux $X$ is
\begin{align}
X= \alpha \,\sigma \wedge \rho + \beta\, \psi~,
\end{align}
where $\alpha$ and $\beta$ are constant parameters and $\psi=\ell^{1234}$, $\sigma=\ell^{56}$ and $\rho=\ell^{78}$. Furthermore
\bea
Q=\gamma_{\tilde a} \bbl^{\tilde a}~,
\eea
where $\gamma$ are constants.  As $Q$ is parallel,
 the field equation for $Q$, \eqref{fqxx}, gives $X\wedge X=0$ and so we obtain the condition that either $\alpha=0$ or $\beta=0$. Let us proceed to investigate $\alpha=0$, as the case for $\beta=0$ can be dealt with in complete analogy. As $X=\beta \psi$, the algebraic KSE \eqref{algKSE} becomes
\begin{align}
({1\over 3A} \slashed {Q}+\frac{\beta}{6a^2} \Gamma^{1234} \Gamma_z )\sigma_+ = \frac{1}{\ell A} \sigma_+~.
\end{align}
The integrability condition of this is
\bea
{1\over 9 A^2} Q^2+\frac{\beta^2}{36a^4}=\frac{1}{\ell^2 A^2}~.
\eea
On the other hand  the warp factor field equation \eqref{warp} gives
\begin{align}
{1\over 6 A^2} Q^2+\frac{\beta^2}{12a^4} = \frac{1}{\ell^2 A^2}~.
\end{align}
The last two equations are incompatible and so there are no supersymmetric solutions.

 \subsubsection{$M^8=S^5\times S^2\times S^1$}

Here we shall consider two cases that  with $S^5=U(3)/U(2), SU(3)/SU(2)$ and that with $S^5=\mathrm{Spin}(6)/\mathrm{Spin}(5)$.  The latter can be excluded immediately.
 As $M^8$ is a product of symmetric spaces all
 left-invariant forms are parallel and represent classes in the de-Rham cohomology of $M^8$.  As  $H^4(S^5\times S^2\times S^1)=0$, we have that $X=0$.
  The solution becomes electric and as we have seen such solutions cannot preserve $N>16$ supersymmetries.

 Next suppose that $S^5=U(3)/U(2)$.
  The metric on $M^8$ can be chosen as
  \bea
  ds^2(M^8)= ds^2(S^5)+ ds^2(S^2)+ ds^2(S^1)~,
  \eea
  where
  \bea
  ds^2(S^5)= b\sum_{r=1}^4 (\bbl^r)^2+ a (\bbl^5)^2~,~~~ds^2(S^2)= c \left((\bbl^6)^2+(\bbl^7)^2\right)~,~~~ ds^2(S^1)= f (\bbl^8)^2~,
  \eea
  and where  $a,b,c,f>0$ are constants.
  The invariant forms are generated by $\bbl^5$, $\bbl^8$, $\omega=\bbl^{12}+ \bbl^{34}$ and  $\sigma= \bbl^{67}$.  The independent differential relations between the invariant forms are
  \bea
  d\bbl^5= \omega~,~~~d\bbl^8=0~,~~~d\sigma=0~,~~~
  \eea
  where we have used the description of the geometry on $S^5$ as in section \ref{u3u2su2}.
  As $dQ=0$, we have that $Q=\gamma \bbl^8$. Furthermore after imposing $dX=0$ the most general flux $X$  is
  \bea
   X={1\over 2} \alpha\, \omega\wedge \omega+\beta\, \omega\wedge \sigma~,
   \label{xu4u3s2s1}
   \eea
   where  $\alpha, \beta$ are constants.

   The algebraic KSE gives
   \bea
  \Big[ {1\over6}\Big( {\alpha\over b^2} \Gamma^{1234}+{\beta\over bc} (\Gamma^{1267}+\Gamma^{3467})\Big)\Gamma_z+{1\over3} {\gamma\over \sqrt f\, A} \Gamma^8\Big]\sigma_+={1\over \ell A} \sigma_+~.
  \label{ksenr8}
  \eea
Squaring this, we find
\bea
\Big[{1\over 36}  \Big( {\alpha^2\over b^4} +{2\beta^2\over b^2c^2} -2{\alpha\beta\over b^3c} (J_1+J_2)+ {2\beta^2\over b^2 c^2} J_1 J_2 \Big)+{1\over9} {\gamma^2\over  f\, A^2} \Big]\sigma_+={1\over \ell^2 A^2} \sigma_+~,
\label{44433c}
\eea
where $J_1=\Gamma^{1267}$ and $J_2=\Gamma^{3467}$.  The decomposition of this condition on $\sigma_+$ into eigenspaces of $J_1$ and $J_2$ is given in table \ref{table43x}.

Each common eigenspace of $J_1$ and $J_2$ has dimension 4.  So to find solutions with $N>16$ supersymmetries, we have to consider at least two of these eigenspaces.  Hence
this would necessarily involve either one of the eigenspinors  $|+,-\rangle$ and $|-,+\rangle$ or both eigenspinors $|\pm,\pm\rangle$.  In the former case taking the difference of the condition
that arises on the fluxes
with  the warp factor field equation
\bea
{1\over12} \Big( {\alpha^2\over b^4} +{2\beta^2\over b^2c^2}\Big)+ {\gamma^2\over 6 f A^2} ={1\over \ell^2 A^2} ~,
\eea
one finds that $\alpha=\beta=\gamma=0$ which is a contradiction.  In the latter case, we find that $\alpha\beta=0$.  Using this and comparing the condition on the fluxes
in table \ref{table43x} with the warp factor field equation above again leads to a contradiction.
There are no AdS$_3$ solutions that preserve $N>16$ supersymmetries with internal space $S^5\times S^2\times S^1$.

\begin{table}[h]
\begin{center}
\vskip 0.3cm
 \caption{Decomposition of (\ref{44433c})  into eigenspaces}
 \vskip 0.3cm

	\begin{tabular}{|c|c|}
		\hline
		$|J_1,J_2\rangle$&  relations for the fluxes\\
		\hline
		$|+,-\rangle$, $|-,+\rangle$& ${1\over 36}   {\alpha^2\over b^4}  +{1\over9} {\gamma^2\over  f\, A^2} ={1\over \ell^2 A^2} $ \\
\hline
		$|\pm,\pm\rangle$& ${1\over 36}  \Big( {\alpha^2\over b^4} +{4\beta^2\over b^2c^2} \mp {4\alpha\beta\over b^3c} \Big)+{1\over9} {\gamma^2\over  f\, A^2} ={1\over \ell^2 A^2}$  \\
		\hline
		\end{tabular}
\vskip 0.2cm
  \label{table43x}
 \end{center}
\end{table}

We have also performed the calculation for $S^5=SU(3)/SU(2)$ which gives rise to an $X$ flux with additional  terms to those in (\ref{xu4u3s2s1}) because of the presence
 of an invariant complex (2,0) form. After some investigation, we find
that again there are no solutions with $N>16$ supersymmetry.

\subsubsection{$SU(3)^{k,l}$}

In this context $SU(3)$ is viewed, up to a discrete identification, as a homogeneous space with isotropy group $SU(2)\times U(1)$ and almost effective transitive group $SU(3)\times SU(2)\times U(1)$, where the inclusion map of $SU(2)\times U(1)$ in $SU(3)\times SU(2)\times U(1)$ is
\bea
(a, z)\rightarrow \left(\begin{pmatrix}a z^k & 0\cr 0 & z^{-2k}\end{pmatrix}, a, z^l\right)
\eea
As we have mentioned the geometry of such cosets is more restrictive than that of $SU(3)$ viewed as the homogeneous space $SU(3)/\{e\}$.
Thus it suffices to investigate whether $SU(3)$ is a solution.  As $SU(3)$ does not admit closed 1-forms, $Q=0$.  In such case $X$ is harmonic. However
$H^4(SU(3), \bR)=0$ and so $X=0$.  This leads to a contradiction as the warp factor field equation cannot be satisfied.

 \subsubsection{$N^{k,l, m}\times S^1= \frac{SU(2) \times SU(3)\times U(1) }{\Delta_{k,l, m}(U(1)^2)\cdot (1\times SU(2))}\times S^1$}

 Let us denote the left-invariant frame along $S^1$ with $\ell^8$, $d\ell^8=0$.  $N^{k,l, m}$ can be thought of as a modification of $N^{k,l}$ and so for the analysis that follows we can use
 the description of the geometry of  $N^{k,l}$ in appendix \ref{appencx2}.  In particular, the most general $Q$ flux is
 \bea
 Q=\gamma_1\, \ell^8+\gamma_2\, \ell^7~.
 \eea
 As $dQ=0$, we deduce that $\gamma_2=0$ and set $\gamma_1=\gamma$.  The most general invariant metric is
 \bea
 ds^2(M^8)=  a\, \left((\ell^5)^2+(\ell^6)^2\right)+ b\, \delta_{rs} ( \ell^r \ell^s+ \hat\ell^r \hat\ell^s)+ c\, (\ell^7)^2+ f\, (\ell^8)^2~,
 \eea
 where $(\ell^r, \hat\ell^r, \ell^5, \ell^6, \ell^7)$, $r,s=1,2$, is a left-invariant frame on $N^{k,l}$, $\ell^8$ is a left-invariant frame on $S^1$ and   $a,b,c,f>0$ are constants.
 Next $X$ can be chosen as
 \bea
 X={1\over2} \alpha_1\, \omega_1\wedge \omega_1+ \alpha_2\, \omega_1\wedge \omega_2+ \alpha_3\, \omega_1\wedge\ell^7\wedge\ell^8+\alpha_4\, \omega_2\wedge\ell^7\wedge\ell^8~,
 \eea
 where $\alpha_1, \alpha_2, \alpha_3, \alpha_4$ are constants. As $dX=0$, one deduces that $\alpha_3=\alpha_4=0$.  Choosing an  orthonormal frame as
 \bea
 &&\bbe^1=\sqrt{a}\, \ell^5~,~~~\bbe^2=\sqrt{a}\, \ell^6~,~~~\bbe^{2r+1}=\sqrt{b}\, \ell^r~,~~~\bbe^{2r+2}=\sqrt{b}\, \hat \ell^r~,~~~
 \cr
 &&\bbe^7=\sqrt{c}\,\ell^7~,~~~\bbe^8=\sqrt{f}\,\ell^8~,
 \eea
 the algebraic KSE can be written as
 \bea
  \Big[ {1\over6}\Big( {\alpha_1\over b^2} \Gamma^{3456}+{\alpha_2\over ab} (\Gamma^{1234}+\Gamma^{1256})\Big)\Gamma_z+{1\over3} {\gamma\over \sqrt f\, A} \Gamma^8\Big]\sigma_+={1\over \ell A} \sigma_+~.
  \label{ksenr8-2}
  \eea
  The form of this KSE is the same as that in (\ref{ksenr8}).  A similar analysis again reveals that there are no solutions that preserve $N>16$ supersymmetries.

\subsection{$N>16$ solutions with $N_R=10$}

The only  superalgebra that gives rise to ten right-handed supersymmetries   is $\mathfrak{osp}(5|2)$ with $(\mathfrak{t}_R)_0=\mathfrak{so}(5)$.
As we are investigating  backgrounds with $N>16$ and we have chosen that $N_L\geq N_R$, we conclude that $10\leq N_L<22$.  Using this and the results of table \ref{table:ads3ksa}, the allowed
 algebras that can act transitively and effectively on the internal spaces are
\bea
&&\mathfrak{so}(n)_L\oplus \mathfrak{so}(5)_R~,~~n=5,6,7,8,9,10~, ~~(N=2n+10)~;~~~
\cr
&&
 \mathfrak{u}(n)_L\oplus \mathfrak{so}(5)_R~,~~n=3,4,5~,~~(N=4n+10)~;~~~
\cr
&&(\mathfrak{sp}(2) \oplus \mathfrak{sp}(1))_L\oplus \mathfrak{so}(5)_R~,~~~(N=26)~;~~~
\cr
&&
\mathfrak{spin}(7)_L\oplus \mathfrak{so}(5)_R~,~~~~(N=26)~;~~~
\cr
&&
(\mathfrak{g}_2)_L\oplus \mathfrak{so}(5)_R~,~~~~(N=24)~.
\label{nr10}
\eea
The only 8-dimensional homogeneous space  that admits such an action by the algebras presented above is
\bea
S^4\times S^4=\mathrm{Spin}(5)/\mathrm{Spin}(4) \times \mathrm{Spin}(5)/\mathrm{Spin}(4)~.
\eea
It remains to examine whether such a background solves the KSE and field equations of 11-dimensional supergravity.

\subsubsection{$S^4 \times S^4$}

The most general invariant metric on $S^4 \times S^4$ is

\begin{align}
ds^2(M^8)= ds^2(S^4)+ds^2(S^4)=a\, \delta_{ij} \ell^i \ell^j + b\, \delta_{rs} \ell^r \ell^s= \delta_{ab} \bbe^a \bbe^b+ \delta_{rs} \bbe^r \bbe^s~,
\end{align}
where $a,b>0$ are constants and $\ell^i$ ($\bbe^i$), $i=1,\dots4$,  and $\ell^r$ ($\bbe^r$), $r=5,\dots, 8$,  are the left-invariant (orthonormal) frames of the two $S^4$'s, respectively. There are no invariant 1-forms on $M^8$, and therefore $Q=0$.
The invariant 4-forms are just the volume forms on the two spheres, hence the most general 4-form flux is
\begin{align}
X = \alpha \bbe^{1234} + \beta \bbe^{5678}~,
\end{align}
where $\alpha,\beta$ are constants.
The field equation  for $Q$, \eqref{fqxx}, yields the condition that either $\alpha =0$ or $\beta=0$. Without loss of generality, we take $\beta=0$. Substituting $X$ into the algebraic KSE \eqref{algKSE}, one finds
\begin{align}
\frac{\alpha}{6} \Gamma^{1234} \Gamma_z \sigma_+ = \frac{1}{\ell A} \sigma_+~,
\end{align}
and  hence obtains
\begin{align}
\frac{\alpha^2}{36 } = \frac{1}{\ell^2 A^2}~,
\end{align}
as an integrability condition. However, the warp factor field equation  \eqref{warp} implies that
\begin{align}
\frac{\alpha^2}{12 }=\frac{1}{\ell^2 A^2}~.
\end{align}
Thus there is a contradiction and there are no such supersymmetric solutions.

\subsection{$N>16$ solutions with $N_R\geq 12$}

Imposing the restriction that  $N_L\geq N_R$, it is easy to see that there are no homogeneous spaces that admit a transitive and effective $\mathfrak{t}_0$ action. This follows from a detailed
examination of the classification results of \cite{castellani}-\cite{bohmkerr} as well as their modifications.

\newsection{$N>16 ~ AdS_3\times_w M^7$ solutions in (massive) IIA}

\subsection{Field equations and Bianchi identities for $N>16$}

 The bosonic fields of (massive) IIA supergravity are the metric $ds^2$, a 4-form field strength $G$, a 3-form field strength $H$, a 2-form field strength $F$, the dilaton $\Phi$  and    the cosmological constant  dressed with the dilaton $S$. Following the description of warped AdS$_3$ backgrounds  in \cite{iiaads}, we write  the fields as
\begin{align}
ds^2 &=2 \bbe^+ \bbe^- + (\bbe^z)^2 +ds^2(M^7)~,\notag\\
G&=\bbe^+\wedge\bbe^-\wedge\bbe^z\wedge Y + X~,\quad H = W \bbe^+\wedge\bbe^-\wedge\bbe^z + Z~,\notag\\
&F~,\quad \Phi~,\quad S~,
\end{align}
where we have used a  null-orthonormal frame $(\bbe^+, \bbe^-, \bbe^i)$, $i=1,\dots, 7$, defined as in \eqref{nullframe} and  $ds^2(M^7) = \delta_{ij} \bbe^i \bbe^j$. The fields $\Phi, W,  S$ and the warp factor $A$ are functions,
$Y$ is a 1-form, $F$ is a 2-form, $Z$ is a 3-form and $X$ is a 4-form on $M^7$.  As the 2-form field strength $F$ is purely magnetic we have denoted the field
and its component on $M^7$ by the same symbol.  This is also the case for $\Phi$ and $S$.  The dependence of the fields on the AdS$_3$ coordinates  is hidden in the definition of the frame $\bbe^+, \bbe^-$ and $\bbe^z$.
The components of the fields in this frame depend only on the coordinates of $M^7$.

As we have demonstrated in  11-dimensional supergravity,  the description for the fields simplifies considerably for AdS$_3$ backgrounds preserving $N>16$ supersymmetries.
In particular, a similar argument to the one presented for 11-dimensional backgrounds gives that the warp factor $A$ is constant. The proof of this is very similar to that given in 11-dimensions
and so we shall not repeat the analysis.  Furthermore it is a consequence of the homogeneity
theorem and Bianchi identities of the theory  that the scalars $\Phi, S$ and $W$ are constant.

To focus the analysis on the IIA AdS$_3$ backgrounds that preserve $N>16$ supersymmetries, we shall impose these conditions on the Bianchi identities, field equations and Killing spinor
equations.  The general formulae can be found in \cite{iiaads}.  In particular taking $A, W, \Phi$ and $S$ to be constant the Bianchi identities can be simplified as
\begin{align}
dZ&=0~,~~~
dF=  SZ~,~~~ S W =0, \quad dX=  Z\wedge F~,\notag\\
dY&=  - W F~.
\label{iiab}
\end{align}
A consequence of this is that either $S=0$ or $W=0$.
Furthermore, the field equations of the form fluxes can be written as
\bea
 &&d*_{{}_7}Z=*_{{}_7}X\wedge F+ S *_{{}_7}F~,~~~d*_{{}_7}F=-W *_{{}_7}Y+ *_{{}_7}X\wedge Z~,~~~d *_{{}_7}Y=-Z\wedge X~,~~~
 \cr
&& d*_{{}_7}X=Z\wedge Y- W X~,
\label{iiaf}
 \eea
 respectively.  As $M^7$ is compact without boundary observe that $d *_{{}_7}Y=-Z\wedge X$ implies that
 \bea
 Z\wedge X=0~.
 \label{zxz}
 \eea
To see this, first observe  that homogeneity  implies that $*_{{}_7}(Z\wedge X)$  is constant. On the other hand   the integral of $Z\wedge X$ over $M^8$ is the constant $*_{{}_7}(Z\wedge X)$ times the volume of $M^8$.  As the integral of $Z\wedge X$ is zero, this constant must vanish giving (\ref{zxz}).

The dilaton field equation is
\begin{align}\label{iiafieldeqs}
 -\frac{1}{12} Z^2+ \frac{1}{2} W^2 +\frac{5}{4} S^2+ \frac{3}{8}F^2 + \frac{1}{96} X^2 - \frac{1}{4}Y^2=0~.
\end{align}
The Einstein equation along $AdS_3$ and $M^7$ implies
\begin{align}\label{einstiia}
& \frac{1}{2} W^2 + \frac{1}{96} X^2 + \frac{1}{4} Y^2 + \frac{1}{4} S^2 + \frac{1}{8} F^2=\frac{2}{\ell^2 A^2}~, \notag\\
&R^{(7)}_{ij} = \frac{1}{12} X^2_{ij} -\frac{1}{2} Y_i Y_j - \frac{1}{96} X^2 \delta_{ij} + \frac{1}{4} Y^2 \delta_{ij} \notag\\
&\quad~~~~~~~~- \frac{1}{4} S^2 \delta_{ij} +\frac{1}{4} Z_{ij}^2 + \frac{1}{2} F^2_{ij} - \frac{1}{8} F^2 \delta_{ij} ~,
\end{align}
where $\nabla$ and $R^{(7)}_{ij}$ denote the Levi-Civita connection and the Ricci tensor of $M^7$, respectively.  The former condition is the warp factor field equation.

\subsection{The Killing spinor equations}

The solutions to the KSEs of (massive) IIA along $AdS_3$ may be written as in  \eqref{ks}, although now $\sigma_\pm$ and $\tau_\pm$ are $\text{Spin}(9,1)$ Majorana spinors which satisfy the lightcone projections $\Gamma_\pm\sigma_\pm=\Gamma_\pm\tau_\pm=0$ and only depend on the coordinates of $M^7$. These are subject to the gravitino KSEs
\begin{align}\label{iiakse}
\nabla^{(\pm)}_i \sigma_\pm = 0~, \quad \nabla^{(\pm)}_i \tau_\pm = 0~,
\end{align}
the dilatino KSEs

\begin{align}\label{iiadilat}
\mathcal{A}^{(\pm)} \sigma_{\pm} = 0~, \quad \mathcal{A}^{(\pm)} \tau_{\pm} = 0~,
\end{align}
and the algebraic KSEs

\begin{align}\label{iiaalgkse}
\Xi^{(\pm)}\sigma_{\pm}=0~, \quad (\Xi^{(\pm)} \pm \frac{1}{\ell})\tau_{\pm}=0~,
\end{align}
where

\begin{align}
\nabla_i^{(\pm)} &= \nabla_i  + \frac{1}{8} \slashed{Z}_i \Gamma_{11} + \frac{1}{8} S \Gamma_i + \frac{1}{16} \sF \Gamma_i \Gamma_{11} + \frac{1}{192} \sX \Gamma_i \pm \frac{1}{8} \sY \Gamma_{zi}~, \notag \\
\mathcal{A}^{(\pm)} &=  \frac{1}{12} \slashed{Z} \Gamma_{11} \mp \frac{1}{2}W \Gamma_z \Gamma_{11} + \frac{5}{4} S + \frac{3}{8} \sF \Gamma_{11} + \frac{1}{96} \sX \pm \frac{1}{4} \sY \Gamma_{z}~, \notag \\
\Xi^{(\pm)} &= \mp\frac{1}{2\ell}  \pm \frac{1}{4} A W \Gamma_{11} - \frac{1}{8} A S \Gamma_z - \frac{1}{16} A \sF \Gamma_z \Gamma_{11} - \frac{1}{192} A \sX \Gamma_z \mp \frac{1}{8} A \sY~.
\end{align}
  If $M^7$ is compact without boundary, one can demonstrate that
\begin{align}
\parallel \sigma_+ \parallel = \text{const}~, \quad \parallel \tau_+ \parallel = \text{const}~, \quad \langle \sigma_+, \tau_+ \rangle =0~,~~~\langle \tau_+, \Gamma_{iz} \sigma_+\rangle=0~.
\end{align}
As in eleven dimensions, the last condition is essential to establish that the warp factor $A$ is constant for IIA AdS$_3$ backgrounds preserving $N>16$ supersymmetries
with compact without boundary internal space $M^7$.

\begin{table}\renewcommand{\arraystretch}{1.3}
	\caption{7-dimensional compact, simply connected,  homogeneous spaces}
	\centering
	\begin{tabular}{c l}
		\hline
		& $M^7=G/H$  \\  
		\hline
		(1)& $\frac{\mathrm{Spin}(8)}{\mathrm{Spin}(7)}= S^7$, symmetric space\\
		(2)&$\frac{\mathrm{Spin}(7)}{G_2}=S^7$ \\
		(3)& $\frac{SU(4)}{SU(3)}$ diffeomorphic to $S^7$ \\
        (4) & $\frac{Sp(2)}{Sp(1)}$ diffeomorphic to $S^7$ \\
		(5) & $\frac{Sp(2)}{Sp(1)_{max}}$, Berger space \\
		(6) & $ \frac{Sp(2)}{\Delta(Sp(1))}=V_2(\bR^5)$  not spin\\
        (7) & $\frac{SU(3)}{\Delta_{k,l}(U(1))}=W^{k,l}$~~ $k, l$ coprime, Aloff-Wallach space\\
		(8)&$\frac{SU(2) \times SU(3) }{\Delta_{k,l}(U(1))\cdot (1\times SU(2))}=N^{k,l}$ ~$k,l$ coprime\\
		(9) & $\frac{SU(2)^3}{\Delta_{p,q,r}(U(1)^2)}=Q^{p,q,r}$ $p, q, r$ coprime\\
(10)&$M^4\times M^3$,~~$M^4=\frac{\mathrm{Spin}(5)}{\mathrm{Spin}(4)}, ~\frac{ SU(3)}{S(U(1)\times U(2))}, ~\frac{SU(2)}{U(1)}\times \frac{SU(2)}{U(1)}$\\
&~~~~~~~~~~~~~~~~$M^3= SU(2)~,~\frac{SU(2)\times SU(2)}{\Delta(SU(2))}$\\
(11)&$M^5\times \frac{SU(2)}{U(1)}$,~~$M^5=\frac{\mathrm{Spin}(6)}{\mathrm{Spin}(5)}, ~\frac{ SU(3)}{SU(2)}, ~\frac{SU(2)\times SU(2)}{\Delta_{k,l}(U(1))},~ \frac{ SU(3)}{SO(3)} $\\
[1ex]
		\hline
	\end{tabular}
	\label{table:nonlin7}
\end{table}

\subsection{$N>16$ solutions with left only supersymmetry}

AdS$_3$ backgrounds admit the same Killing  superalgebras in  11-dimensional, IIA and IIB supergravities.  As a result the
  Lie algebras $\mathfrak{t}_0$  that must act transitively  and effectively on the
internal spaces of IIA and IIB  AdS$_3$ backgrounds can be read off  those  found in the  11-dimensional  analysis.  So for $N_R=0$, these are given in (\ref{nr0}).  An inspection of the 7-dimensional homogeneous spaces in table \ref{table:nonlin7}   reveals that there are no $N>16$ supersymmetric AdS$_3$ backgrounds with $N_R=0$.

\subsection{$N>16$ solutions with $N_R=2$}

The 7-dimensional homogeneous spaces\footnote{There are several embeddings of $Sp(1)$ in $Sp(2)$ however only one of them admits a modification such that
 the internal space is associated to a background that can preserve $N>16$ supersymmetries.} that admit an effective and transitive action of the Lie algebras in (\ref{nr2}) are
\bea
&&S^7=\mathrm{Spin}(8)/\mathrm{Spin}(7)~~(N=18)~,~~~
\cr
&&S^7=U(4)/U(3)~,~~~(N=18)~,
\cr
&&S^7=(Sp(2)\times Sp(1))/Sp(1)\times Sp(1)~,~~~(N=18)~,
\cr
&& S^4\times S^3=\mathrm{Spin}(5)/\mathrm{Spin}(4)\times SU(2)~~(N=18)~,~~~
\cr
&& S^7=\mathrm{Spin}(7)/G_2~~(N=18)~.
\label{cnr2}
\eea
Solutions with internal space $\mathrm{Spin}(8)/\mathrm{Spin}(7)$ can be immediately excluded.  This is a symmetric space and so all fluxes are parallel.  On the other hand
  the only parallel forms on $S^7$ are the constant
functions and the volume form.  Therefore all k-form fluxes for $k>0$ must vanish.  In such a case the dilaton field equation in (\ref{iiafieldeqs}) implies that $W=S=0$.  In turn, the warp
factor field equation in (\ref{einstiia})  becomes inconsistent.  The remaining cases are investigated below.

\subsubsection{$S^7=U(4)/U(3)$} \label{u4u3x}

The geometry of $S^7=U(4)/U(3)$ has been summarized in the beginning of section \ref{s7u4u3}. The metric is given in (\ref{metru4u3}).  The invariant forms
are generated by the 1-form $\ell^7$ and 2-form $\omega$ as in (\ref{omegau4u3}), $d\ell^7=\omega$.
Given these data, the most general invariant  fluxes can be chosen as
\bea
X={\alpha\over2}\, \omega^2~,~~~Z=\beta\, \ell^7\wedge \omega~,~~~F=\gamma\, \omega~,~~~Y=\delta\, \ell^7~.
\eea
As the Bianchi identities require that $dZ=0$, we have $\beta=0$.  Furthermore the remaining Bianchi identities imply
\bea
SW=0~,~~~\delta=-W\gamma~,
\eea
and the field equations for the fluxes give
\bea
{\alpha\,\gamma \over b}+{1\over2} \gamma\, S\, b=0~,~~~\gamma\, \sqrt{a}=-{1\over3}W{b^2\delta\over \sqrt{a}}~,~~~{\alpha\sqrt{a}\over b}=-{1\over2} W\,\alpha~.
\eea

Suppose first that $S\not=0$. Then $W=0$ which in turn gives $\alpha=\gamma=\delta=0$.  As both $Z=Y=0$, the dilaton field equation in (\ref{iiafieldeqs}) implies that the rest of the fluxes vanish which in turn
leads to a contradiction as the warp factor field equation in (\ref{einstiia})  cannot be satisfied.

Next suppose that $S=0$. Then $\alpha\,\gamma=0$.  Take that $W\not=0$ otherwise there will be a contradiction as described for $S\not=0$ above.  If $\gamma=0$, this will imply that $\delta=0$
and so again the dilaton field equation in (\ref{iiafieldeqs}) will imply that the rest of the fluxes must vanish.

It remains to investigate the case  $\alpha=0$.  The dilatino KSE (\ref{iiadilat}) and algebraic KSE (\ref{iiaalgkse}) become
\bea
&&(-{1\over2} W \Gamma_{11}+{3\over8} \slashed{F} \Gamma_z \Gamma_{11}-{1\over4} \slashed{Y})\sigma_+=0~,
\cr
&&({1\over2} W \Gamma_{11}-{1\over8} \slashed{F} \Gamma_z \Gamma_{11}-{1\over4} \slashed{Y})\sigma_+={1\over \ell A} \sigma_+~.
\label{wfyalg}
\eea
Eliminating the flux $F$, one finds
\bea
(W \Gamma_{11}-\slashed{Y})\sigma_+={3\over\ell A}\sigma_+~.
\label{wyalg}
\eea
The integrability condition gives
\bea
W^2+ Y^2={9\over\ell^2 A^2}~.
\label{wfyint}
\eea
Comparing this with the field equation for the warp factor (\ref{einstiia}) leads to a contradiction.  There are no supersymmetric solutions.

\subsubsection{$S^7=(Sp(2)\times Sp(1))/Sp(1)\times Sp(1)$}

The geometry of $S^7=(Sp(2)\times Sp(1))/Sp(1)\times Sp(1)$  has been described in section \ref{sp2sp1sp1sp1s1}. As it has been explained
there are no invariant 1- and 2-forms, and no invariant closed 3-forms on this homogeneous space.  As a result $Y=F=Z=0$.  Then the dilaton field equation in (\ref{iiafieldeqs}) implies that
$W=S=X=0$ and therefore the warp factor field equation  in (\ref{einstiia}) cannot be satisfied.  There are no AdS$_3$ solutions with internal space $S^7=(Sp(2)\times Sp(1))/Sp(1)\times Sp(1)$.

\subsubsection{$M^7=S^4\times S^3=\mathrm{Spin}(5)/\mathrm{Spin}(4)\times SU(2)$} \label{o5o4su2}

The metric on the internal space can be chosen as
\bea
ds^2(M^7)=  ds^2(S^4)+ ds^2(S^3)= a\, \delta_{ij} \bbl^i\bbl^j+ b_{rs} \bbl^r \bbl^s~,~~~
\eea
where $(\bbl^i)$, $i=4,\dots, 7$,  is a left-invariant frame on $S^4$ and $(\bbl^r)$, $r=1,2,3$ is a left-invariant frame on $S^3=SU(2)$,
$a>0$ is a constant and $(b_{rs})$ a positive definite $3\times 3$ symmetric matrix. Note that
\bea
d\bbl^r={1\over2} \epsilon^r{}_{st} \bbl^s\wedge \bbl^t~.
\eea
Before we proceed observe that without loss of generality $b=(b_{rs})$ can be chosen  to be
diagonal.  This is because  any transformation $\bbl^r\rightarrow O^r{}_s \bbl^s$ of the left-invariant frame with $O\in SO(3)$ leaves the structure constants of $\mathfrak{su}(2)$
invariant and acts on $b$ as $O^t b O$.  So there is a choice of frame such that $b=\mathrm{diag}(b_1, b_2, b_3)$ with $b_1,b_2,b_3>0$ constants.  From here on we shall take
$b$ to be diagonal.

The most general invariant fluxes are
\bea
X=\alpha\, \bbl^{4567}~,~~~Z=\beta\, \bbl^{123}~,~~~F={1\over2}\gamma_r\, \epsilon^r{}_{st} \bbl^s\wedge \bbl^t~,~~~Y=\delta_r \bbl^r~,
\eea
where $\alpha$, $\beta$, $\gamma_r$ and $\delta_r$, $r=1,2,3$, are constants.

First observe that $Z\wedge X=0$ implies that $\alpha \beta=0$.   Next suppose that $S\not=0$.  It follows from the Bianchi identities (\ref{iiab}) that  $Z=0$  as $dF=0$.  In addition,    the Bianchi identities (\ref{iiab}) give
\bea
W=Y=0~.
\eea
Next the dilaton field equation in (\ref{iiafieldeqs}) implies that $S=X=F=0$ which is a contradiction to the assumption that $S\not=0$.

So let us now consider that $S=0$.  Again $\alpha \beta=0$ and so either $Z=0$ or $X=0$.  Let us first take $Z=0$ and $X\not=0$.  In such a case the field equation for $X$ in (\ref{iiaf})
gives $W=0$.  If $W=0$, the Bianchi identities (\ref{iiab}) will imply that $Y=0$.  This in turn leads to a contradiction as the field equation for the dilaton in (\ref{iiafieldeqs}) implies that
$X=0$.

Suppose now that both $Z=X=0$. As $S=0$ as well,  the dilatino and algebraic KSEs can be re-written as in (\ref{wfyalg}).  This in turn gives (\ref{wyalg}) that leads to
the integrability condition (\ref{wfyint}).
Substituting this into the field equation for the warp factor in (\ref{einstiia})  and after  eliminating $Y^2$, one finds a contradiction.

It remains to investigate the case that $Z\not=0$ and $X=0$.  First the Bianchi identity for $Y$  (\ref{iiab}) implies that
\bea
\delta_r=-W \gamma_r~.
\label{dwg}
\eea
Then field equation for $F$, $d*_{{}_7} F=-W *_{{}_7} Y$, together with (\ref{dwg}) imply that
\bea
W^2 \,\gamma_r={b_r^2\over b_1 b_2 b_3}\, \gamma_r~,~~~\text{no summation over $r$}~.
\label{w2b3}
\eea
Next turn to the Einstein equation along  $S^4$.  As $X=0$ and the fields $Z$, $F$ and $Y$ have non-vanishing components only along $S^3$, we find that
\bea
R^{(7)}_{ij}= ({1\over 4} Y^2-{1\over 8} F^2) \delta_{ij}~.
\eea
Using (\ref{w2b3}), one can show that $R^{(7)}_{ij}=0$.  This   is a contradiction as $R^{(7)}_{ij}$ is the Ricci tensor of the  $S^4$ subspace which is required to be  strictly positive.  Therefore we conclude
that there are no supersymmetric IIA AdS$_3$ solutions with internal space $S^4\times S^3$.

\subsubsection{$M^7=S^7=\mathrm{Spin}(7)/G_2$}
This homogeneous space admits invariant 3- and 4-forms which are the fundamental  $G_2$ forms  $\varphi$ and $*{{}_7}\varphi$.  However the 3-form $\varphi$ is not closed and so $Z=0$.
As there are no invariant 1-forms and 2-forms $Y=F=0$.  In such a case the dilaton field equation in (\ref{iiafieldeqs})  implies that $W=S=X=0$.  In turn, the warp factor field equation in
(\ref{einstiia}) becomes inconsistent.

\subsection{$N>16$ solutions with $N_R=4$}

The Lie algebras that must act both effectively and transitively on the internal space $M^7$ are the same as those found in $D=11$ supergravity and  given in (\ref{nr4}).  After an inspection of the 7-dimensional homogeneous
spaces those admitting an effective and transitive action by such Lie algebras are the following
\bea
S^6\times S^1= \mathrm{Spin}(7)/\mathrm{Spin}(6)\times S^1~,~~(N=18)~; ~G_2/SU(3)\times S^1~,~~(N=18)~.
\label{cnr4}
\eea
Both are products of 6-dimensional homogeneous spaces with $S^1$.

\subsubsection{$M^7=\mathrm{Spin}(7)/\mathrm{Spin}(6)\times S^1$}

The only non-vanishing k-form flux, $k>0$, allowed is $Y=\alpha \bbl^7$, where $\bbl^7$ is a left-invariant frame along $S^1$ and $\alpha$ is constant.  The dilatino
KSE (\ref{iiadilat}) can be re-written as
\bea
\Big(-{1\over2} W\Gamma_{11}+{5\over4} S\Gamma_z-{1\over4} \slashed{Y}\Big)\sigma_+=0~,
\eea
which leads to the integrability condition
\bea
W^2+{25\over 4}S^2+{1\over 4} Y^2=0~.
\eea
As a result $W=S=Y=0$.  This leads to an inconsistency as the warp field equation (\ref{einstiia}) cannot be satisfied.
There are no supersymmetric IIA AdS$_3$ backgrounds with $\mathrm{Spin}(7)/\mathrm{Spin}(6)\times S^1$ internal space.

\subsubsection{$M^7=S^6\times S^1=G_2/SU(3)\times S^1$} \label{g2su3s1}

The differential algebra of a  left-invariant frame on $M^7$ modulo terms in $\mathfrak{su}(3)\wedge \mathfrak{m}$ which involve the canonical connection is
\bea
d\lambda^{\bar r}={1\over2} \epsilon^{\bar r}{}_{st} \lambda^r\wedge \lambda^t~,~~~d\bbl^7=0~,~~~r=1,2,3~,
\eea
where $\lambda^r$ is a complex frame, $\bar{\lambda}^r= \lambda^{\bar r}$, on $S^6$ and $\bbl^7$ is a left-invariant frame on $S^1$.  The invariant forms on $S^6$ are the 2-form
\bea
\omega={i\over 2} \delta_{r\bar s} \lambda^r\wedge \lambda^{\bar s}~,
\eea
and the holomorphic 3-form
\bea
\chi={1\over 6} \epsilon_{rst} \lambda^r\wedge \lambda^s\wedge \lambda^t~.
\eea
Clearly
\bea
d\omega=3\,\mathrm{Im} \chi~,~~~d\,\mathrm{Re}\chi=2\omega\wedge \omega~.
\eea
The most general invariant metric  on $M^7$ is
\bea
ds^2(M^7)=a\, \delta_{r\bar s} \lambda^r \lambda^{\bar s}+ b\, (\bbl^7)^2~,
\eea
where $a,b>0$ are constants. Moreover the most general invariant fluxes are
\bea
X= {1\over 2}\alpha_1\,\omega^2+\alpha_2 \bbl^7\wedge \mathrm{Re}\,\chi+\alpha_3 \bbl^7\wedge \mathrm{Im}\,\chi~,~~~Z=\beta\, \mathrm{Im} \chi~,~~~F=\gamma\, \omega~,~~~Y=\delta\, \bbl^7~,
\eea
where $\alpha$'s, $\beta, \gamma$ and $\delta$ are constants and we have used that $dZ=0$.  As $Z\wedge X=0$, we have that $\alpha_2 \beta=0$.
Furthermore as $dF=SZ$, we have that
\bea
3\,\gamma=S\beta~.
\eea

Let us first consider the case that $S\not=0$.  This implies that $W=0$.  As either $\alpha_2=0$ or $Z=0$, let us investigate first the case that $Z=0$.  In such a case  the Bianchi identities
(\ref{iiab}) and the field equations (\ref{iiaf}) imply that $X$ is harmonic and as $H^4(S^6\times S^1)=0$, we have $X=0$. Using this, we also find  that  $F$ is harmonic and so as $H^2(S^6\times S^1)=0$, $F=0$.  Next the dilatino KSE (\ref{iiadilat}) becomes $(5S+\slashed{Y}\Gamma_z)\sigma_+=0$ which in turn implies that $25S^2+Y^2=0$.  This is a contradiction as $S=0$.
Thus there are no such supersymmetric AdS$_3$ backgrounds.

Next suppose that $\alpha_2=0$. The field equation $d*_{{}_7}X=Z\wedge Y$ gives
\bea
\alpha_3=0~,
\eea
and so $X$ does not have a component along $S^1$.  Then the Einstein equation along $S^1$ gives
\bea
R^{(7)}_{77}=-{1\over4} Y^2-{1\over 4} S^2-{1\over 8} F^2-{1\over 96} X^2=0~.
\eea
Thus again $S=0$ which is a contradiction.

So to find solutions, we have to set $S=0$.  The Bianchi identity $dF=0$ gives $F=0$.   Furthermore the field equation $d*_{{}_7}Z=0$ gives $Z=0$.
We also have from the field equations (\ref{iiaf}) that $W *_{{}_7}Y=0$.  If we choose $Y=0$ and as $Z$  vanishes as well, $Z=0$ ,  the dilaton field equation in (\ref{iiafieldeqs}) implies that the rest
of the fields vanish which contradicts the warp factor field equation. So let us take $W=0$.  In such a case $X$ is harmonic and so $X=0$. This is also the case for $Z$ and so $Z=0$. In turn the dilatino KSE
implies that $\slashed{Y}\sigma_+=0$ which gives $Y=0$.  Thus all the fields vanish leading to a contradiction with the warp factor field equation.  There are no  supersymmetric AdS$_3$ solutions
with internal space $G_2/SU(3)\times S^1$.

\subsection{$N>16$ solutions with $N_R=6$}
The 7-dimensional homogeneous spaces that admit an effective and transitive action of one of the Lie algebras in (\ref{nr6}) are
\bea
&&S^5\times S^2=\mathrm{Spin}(6)/\mathrm{Spin}(5)\times SU(2)/U(1)~,~~(N=18)~;
\cr
&&S^5\times S^2=U(3)/U(2)\times SU(2)/U(1)~,~~(N=18)~;
\cr
&&S^4\times S^3=\mathrm{Spin}(5)/\mathrm{Spin}(4)\times \big(SU(2)\times SU(2)\big)/SU(2)~,~~(N=22)~.
\label{cnr6}
\eea
The $\mathrm{Spin}(6)/\mathrm{Spin}(5)\times SU(2)/U(1)$ case can be easily ruled out as $Y=Z=0$.  Then the dilaton field equation implies that $X=W=S=F=0$
which in turn leads to a contradiction as the warp factor field equation cannot be satisfied. Moreover the $\mathrm{Spin}(5)/\mathrm{Spin}(4)\times \big(SU(2)\times SU(2)\big)/SU(2)$
internal space has been investigated already as it is a special case of $\mathrm{Spin}(5)/\mathrm{Spin}(4)\times SU(2)$.

\subsubsection{$S^5\times S^2=U(3)/U(2)\times SU(2)/U(1)$}\label{u3u2su2u1}

The geometry of the homogeneous space  $S^5=U(3)/U(2)$ has been described in section \ref{u3u2su2}.  Using this, the most general invariant metric on $M^7$ can be written as

\bea
ds^2(M^7)= ds^2(S^5)+ds^2(S^2)= a\, (\ell^5)^2+ b\, \delta_{rs} \ell^r \ell^s+ c\,\left( (\ell^6)^2+(\ell^7)^2\right)~,
\label{metru3u2su2u1}
\eea
where $a,b,c>0$ are constants $(\ell^r, \ell^5)$, $r=1,2,3,4$,  is a left-invariant frame on $S^5$ and $(\ell^6, \ell^7)$ is a left-invariant frame
on $S^2$.  The invariant forms on the homogeneous space are generated by
\bea
\bbl^5~,~~~\omega=\ell^{12}+\ell^{34}~,~~~\sigma=\ell^{67}~,
\eea
with
\bea
d\bbl^5=\omega~,~~~d\sigma=0~.
\eea
The most general invariant fluxes are
\bea
&&X={1\over2} \alpha_1\, \omega^2+\alpha_2\, \omega\wedge \sigma~,~~~Z=\beta_1\, \ell^5\wedge \omega+ \beta_2\, \ell^5\wedge \sigma~,
\cr
&& F=\gamma_1\, \omega+\gamma_2 \sigma~,~~~Y=\delta\, \ell^5~.
\eea
The Bianchi identity $dZ=0$ implies that $\beta_1=\beta_2=0$.  So $Z=0$.  The remaining Bianchi identities imply that
\bea
SW=0~,~~~\delta=-W\gamma_1~,~~~W \gamma_2=0~.
\label{xxe}
\eea

To continue first take $S\not=0$.  In such a case $W=0$ and so $\delta=0$.  As both $Y=Z=0$, the dilaton field equation implies that $S=X=F=0$.  This is a contradiction to the assumption that $S\not=0$.

Therefore we have to set $S=0$. Furthermore $W\not=0$ as otherwise $Y=Z=0$ and the dilaton field equation will imply that all other fluxes must vanish.  This in turn leads to a contradiction as the warp factor field equation cannot be satisfied. As $W\not=0$, we have $\gamma_2=0$.  Then  the field equation for the fluxes (\ref{iiaf})  give $*_{{}_7}X\wedge F=0$ which in turn implies that
\bea
\alpha_2 \gamma_1=0~,~~~\gamma_1\alpha_1=0~.
\eea
Notice that $\gamma_1\not=0$ as otherwise $\delta=0$ and so $Y=Z=0$ leading again to a contradiction. Thus we find $\alpha_1=\alpha_2=0$ and so $X=0$.

As we have established that $Z=X=0$, we can follow the analysis of the KSEs in section \ref{u4u3x} that leads to the conclusion
that there are no supersymmetric AdS$_3$ solutions with internal space $U(3)/U(2)\times SU(2)/U(1)$.

\subsection{$N>16$ solutions with $N_R=8$}
The 7-dimensional homogeneous spaces that admit an effective and transitive action of one of the Lie algebras in (\ref{nr8}) are
\bea
&&S^7=(Sp(2)\times Sp(1))/Sp(1)\times Sp(1)~,~~~(N=18)~;
\cr
&&S^4\times S^3=\mathrm{Spin}(5)/\mathrm{Spin}(4)\times SU(2)~,~~~(N=18)~;
\cr
&&S^4\times S^2\times S^1=\mathrm{Spin}(5)/\mathrm{Spin}(4)\times SU(2)/U(1)\times S^1~,~~~(N=18)~;
\cr
&&S^5\times S^2=\mathrm{Spin}(6)/\mathrm{Spin}(5)\times SU(2)/U(1)~,~~~(N=20)~;
\cr
&&S^5\times S^2=U(3)/U(2)\times SU(2)/U(1)~,~~~(N=20)~;
\cr
&& N^{k,l,m}=(SU(2)\times SU(3)\times U(1))/\Delta_{k,l,m} (U(1)\times U(1))\cdot (1\times SU(2))~,~~~(N=20)~;
\cr
&&S^4\times S^3=\mathrm{Spin}(5)/\mathrm{Spin}(4)\times \big(SU(2)\times SU(2)\big)/SU(2)~,~~~(N=18)~.
\label{cnr8}
\eea
The only new cases that arise and has not been investigated already are those with internal space $S^4\times S^2\times S^1=\mathrm{Spin}(5)/\mathrm{Spin}(4)\times SU(2)/U(1)\times S^1$
and $N^{k,l,m}$.
All the remaining ones do not give supersymmetric solutions with $N>16$ and $N_R=8$.

\subsubsection{$S^4\times S^2\times S^1=\mathrm{Spin}(5)/\mathrm{Spin}(4)\times SU(2)/U(1)\times S^1$} \label{so5so4su2u1s1}

The most general invariant  metric on this homogeneous space can be written as
\bea
ds^2(M^7)= ds^2(S^4)+  ds^2(S^2)+ ds^2(S^1)=a\, \delta_{rs} \ell^r \ell^s+ b\, \big((\ell^5)^2+ (\ell^6)^2\big)+ c\, (\ell^7)^2~,
\label{metrso5so4su2u1s1}
\eea
where $a,b,c>0$ are constants and $\ell^r$, $r=1,2,3,4$, is a left-invariant frame on $S^4$, $(\ell^5, \ell^6)$ is a left-invariant frame on $S^2$ and $\ell^7$
is a left-invariant frame on $S^1$.  The most general invariant form fluxes can be chosen as
\bea
X=\alpha\, \ell^{1234}~,~~~Z= \gamma\, \ell^7\wedge \ell^{56}~,~~~F=\beta\, \ell^{56}~,~~~Y=\delta\, \ell^7~,
\eea
where $\alpha, \beta, \gamma, \delta$ are constants.

From the Bianchi identities (\ref{iiab}) and  the field equation (\ref{zxz}), we find that
\bea
S\gamma=0~,~~~\alpha\gamma=0~,~~~SW=0~,~~~W\beta=0~.
\eea
First suppose that $S\not=0$.  It follows that $W=Z=0$.  Moreover from the field equation of $Z$ (\ref{iiaf}) follows that $F=0$.  Next consider the Einstein field equation
to find that
\bea
R^{(7)}_{77} &= -\frac{1}{4 } Y^2 - \frac{1}{96} X^2 - \frac{1}{4} S^2~.
\eea
However this is the Ricci tensor of $S^1$ and hence vanishes. This in turn gives $S=0$ which is a contradiction to our assumption that $S\not=0$.

Thus we have to set $S=0$. The Bianchi identities (\ref{iiab}) and  the field equation (\ref{zxz}) give that
\bea
\alpha \gamma=0~,~~~W\beta=0~,
\eea
and the field equations (\ref{iiaf}) of the form field strengths imply that
\bea
W \delta=0~,~~~W\alpha=0~.
\eea
Therefore if $W\not=0$, we will have $F=Y=X=0$.  Furthermore as $R^{(7)}_{77}=0$, the Einstein equation reveals that $Z=0$.
 Then the dilaton field equation implies that $W=0$ which is a contradiction to our assumption that $W\not=0$.

It remains to investigate solutions with $W=S=0$.  Notice that we should have that $Z\not=0$, or equivalently $\gamma\not=0$, as otherwise the Einstein equation
$R^{(7)}_{77}=0$  will imply that $X=Y=F=0$ and so the warp factor field equation cannot be satisfied leading to a contradiction.
Thus $Z\not=0$ and as $\alpha\gamma=0$, we have that $X=0$.  Inserting $X=S=W=0$ into the dilatino and algebraic KSEs we find that they can be rewritten as
\bea
&&\Big(-{\gamma\over b\sqrt c} J_1 J_2+{3\over2} {\beta\over b} J_1-{1\over2} {\delta\over \sqrt c} J_2\Big)\sigma_+=0~,
\cr
&&\Big({\beta\over b} J_1+{\delta\over \sqrt c} J_2\Big)\sigma_+=-{4\over \ell A} \sigma_+~,
\label{systkse}
\eea
where $J_1=\Gamma^{56} \Gamma_z \Gamma_{11}$ and $J_2=\Gamma^7$.  As each common eigenspace of  $J_1$ and $J_2$ has dimension 4 to find solutions with $N>16$ supersymmetries we have to choose
at least two of these eigenspaces.  One can after some calculation verify that for all possible pairs of eigenspaces the resulting system of equations arising from (\ref{systkse})
does not have solutions.  Therefore there are no AdS$_3$ solutions that have internal space $\mathrm{Spin}(5)/\mathrm{Spin}(4)\times SU(2)/U(1)\times S^1$ and preserve
$N>16$ supersymmetries.

\subsubsection{$N^{k,l,m}=(SU(2)\times SU(3)\times U(1))/\Delta_{k,l,m} (U(1)\times U(1))\cdot (1\times SU(2))$}
As $N^{k,l,m}$ is a modification of $N^{k,l}$, see \cite{witten, fre}, we can use the local description of the geometry of the latter in appendix \ref{appencx2} to describe the former.
In particular the metric can be written as
\bea
ds^2(M^7)= a\, (\ell^7)^2+ b\, (\delta_{rs} \ell^r \ell^s+\delta_{rs} \hat\ell^r \hat\ell^s)+ c\, ((\ell^5)^2+(\ell^6)^2)~,~~~r,s=1,2~,
\label{metrnklm}
\eea
where $(\ell^r, \hat\ell^r, \ell^5, \ell^6, \ell^7)$ is a left-invariant frame and $a,b,c>0$ constants. From the results of appendix \ref{appencx2}, one can deduce that there
are no closed 3-forms and so $Z=0$. The remaining invariant form field strengths are
\bea
X={1\over2}\alpha_1\omega_1^2+\alpha_2 \omega_1\wedge \omega_2~,~~~F=\gamma_1\omega_1+\gamma_2 \omega_2~,~~~Y=\delta \ell^7~,
\eea
where $\alpha_1, \alpha_2, \gamma_1, \gamma_2, \delta$ are constants.
The Bianchi identities (\ref{iiab})  imply that
\bea
SW=0~,~~~-{\delta\over 8 l}=\gamma_1 W~,~~~{\delta\over 4k}=\gamma_2 W~.
\label{bbb5}
\eea
Furthermore, the field equation for $Z$ in (\ref{iiaf}) gives
\bea
&&{c\over b^2} \alpha_1 \gamma_1+ {1\over c} \alpha_2 \gamma_2+ S c \gamma_1=0
\cr
&& \alpha_2 \gamma_1+{1\over2} S \gamma_2 b^2=0~.
\label{fff5}
\eea
Clearly from (\ref{bbb5}) either $S=0$ or $W=0$.  Suppose that $S\not=0$.  Then $W=0$ and from the rest of the conditions arising in the Bianchi identities $Y=0$.
As both $Y=Z=0$, the dilaton field equation implies that $S=F=X=W=0$ which is  a contradiction to our assumption that $S\not=0$.

Therefore we set $S=0$.  We also take $W\not=0$ as otherwise the same argument presented above leads again into a contradiction.
As $S=0$, the last condition in (\ref{fff5}) implies that $\alpha_2 \gamma_1=0$.  However $\gamma_1$ cannot vanish. Indeed if $\gamma_1=0$, then (\ref{bbb5}) will  lead to $Y=0$. Since $Y=Z=0$, the dilaton field equation in (\ref{iiafieldeqs})  will imply that the rest of the fields vanish.  In turn the warp factor field equation (\ref{einstiia}) cannot be satisfied.  Thus we have to set $\gamma_1\not=0$.  In such case  $\alpha_2=0$ and the first equation in (\ref{fff5}) gives $\alpha_1=0$.  As both $\alpha_1=\alpha_2=0$, $X=0$.

We have shown that the remaining non-vanishing fields are $W$, $Y$ and $F$.  To continue consider the dilatino and algebraic KSEs. These can be written as in (\ref{wfyalg}).
Then  a similar argument as that presented in section \ref{u4u3x} leads to a contradiction.  There are no supersymmetric AdS$_3$ solutions with internal space $N^{k,l,m}$.

\newsection{$N>16 ~ AdS_3\times_w M^7$ solutions in IIB}

\subsection{Field equations and Bianchi identities for $N>16$}

The bosonic fields of IIB supergravity are a metric $ds^2$, a complex 1-form field strength $P$, a complex 3-form field strength $G$ and a real self-dual 5-form field strength $F$.
For the investigation of IIB $AdS_3\times_w M^7$ backgrounds that  follows, we shall use the analysis presented  in \cite{iibads} where all the necessary  formulae can be found.  As we are focusing on backgrounds that preserve $N>16$ supersymmetries,
the homogeneity theorem implies that the scalars are constants and so $P=0$.  We shall use this from the beginning to simplify the relevant field equations, Bianchi identities and KSEs.
Imposing the symmetries of the AdS$_3$ subspace on the fields, one finds that the non-vanishing fields are
\begin{align}
ds^2 &= 2 \bbe^+ \bbe^- + (\bbe^z)^2 +ds^2(M^7)~, \quad F= \bbe^+\wedge \bbe^- \wedge \bbe^z\wedge Y - *_{{}_7} Y \notag\\
G &= X\, \bbe^+\wedge\bbe^-\wedge \bbe^z + H~, \
\end{align}
where a null-orthonormal frame $(\bbe^+, \bbe^-, \bbe^z, \bbe^i)$, $i=1,\dots, 7$,  is defined as \eqref{nullframe} and  $ds^2(M^7) = \delta_{ij} \bbe^i \bbe^j$. $Y$ is a real 2-form, $X$ is a complex function and $H$ a complex 3-form on $M^7$. The dependence of the fields on AdS$_3$ coordinates is hidden in the definition of the frame $(\bbe^+, \bbe^-, \bbe^z)$.  All the components
of the fields in this frame depend on the coordinates of $M^7$.

The Bianchi identities of the k-form  field strengths can be written as
\begin{align}
dY&= \frac{i}{8} (\overline{X} H- X \overline{H})~,\quad dX= 0 \notag\\
d*_7Y&= -\frac{i}{8} H\wedge\overline{H}~, \quad dH= 0~, \label{iibb}
\end{align}
while their field equations are
\bea\label{iibf}
 \frac{1}{6} H^2 +  X^2=0~,~~~
 d*_{{}_7} H=4i X *_{{}_7}Y+ 4i Y\wedge H~.
\eea
Note that the Bianchi identities imply that $X$ is constant.  We have also used that the warp factor $A$ is constant.  This is proved as in eleven dimensions upon making use
of the compactness of $M^7$ and the homogeneity theorem.

The Einstein equation along AdS$_3$ and $M^7$ becomes
\begin{align}\label{iibeinst}
 &2Y^2 +\frac{3}{8} {X \overline{X}} + \frac{1}{48} {H_{ijk} \overline{H}^{ijk}} = \frac{2}{\ell^2} A^{-2}~, \notag\\
R_{ij}^{(7)}&=  2 Y^2 \delta_{ij} - 8 Y^2_{ij} + \frac{1}{4} H_{(i}{}^{kl} \overline{H}_{j)kl} \notag\\
&\quad + \frac{1}{8} {X \overline{X}} \delta_{ij} - \frac{1}{48} {H_{klm} \overline{H}^{klm}} \delta_{ij}~,
\end{align}
respectively. Here, $\nabla$ denotes the Levi-Civita connection on $M^7$ and $R^{(7)}$ is the Ricci tensor on the transverse space.  The first condition above is
the field equation for the warp factor.

\subsection{The Killing spinor equations}

The solution of the KSEs of IIB supergravity along the $AdS_3$-subspace can be expressed as in \eqref{ks}, only that now $\sigma_\pm$ and $\tau_\pm$ are $\text{Spin}(9,1)$  Weyl spinors which depend only on the coordinates of $M^7$ and satisfy the lightcone projections $\Gamma_\pm \sigma_{\pm}=\Gamma_\pm \tau_\pm =0$. The remaining independent KSEs are the gravitino
\begin{align} \label{iibgravkse}
\nabla^{(\pm)}_i \sigma_\pm = 0~, \quad \nabla^{(\pm)}_i \tau_\pm = 0~,
\end{align}
dilatino
\begin{align}\label{algkseiib2}
\mathcal{A}^{(\pm)} \sigma_\pm = 0~, \quad \mathcal{A}^{(\pm)} \tau_\pm = 0~,
\end{align}
and algebraic
\begin{align} \label{algkseiib}
\Xi^{(\pm)} \sigma_\pm = 0~, \quad \left(\Xi^{(\pm)} \pm \frac{1}{\ell} \right) \tau_\pm = 0~,
\end{align}
KSEs, where
\begin{align}
\nabla^{(\pm)}_i &= \nabla_i \pm \frac{i}{4} \sgY_i \Gamma_z \mp \frac{i}{2} \sY_i \Gamma_z \notag\\
&\quad +\left(-\frac{1}{96} \sgH_i + \frac{3}{32} \sH_i \mp \frac{1}{16} X \Gamma_{zi}\right)C*~,\notag\\
\mathcal{A}^{(\pm)} &= \mp\frac{1}{4} X \Gamma_z + \frac{1}{24} \sH~,\notag\\
\Xi^{(\pm)} &= \mp\frac{1}{2\ell}  \pm \frac{i}{4} A \sY + \left(\frac{1}{96} A \Gamma_z \sH \pm \frac{3}{16} A X\right)C*~,
\end{align}
and $C$ is the charge conjugation matrix followed by complex conjugation. In the expressions above we have used that $P=0$ and that $A$ is constant.
As in 11-dimensional and IIA supergravities, the IIB AdS$_3$  backgrounds preserve an even number of supersymmetries.

\subsection{$N>16$ solutions with $N_R=0$ and  $N_R=2$}

The existence of solutions that preserve strictly 28 and 30 supersymmetries has already been excluded in \cite{iibn28}.
As in the IIA case,  IIB $N>16$ supersymmetric AdS$_3$ solutions with $N_R=0$ can also be ruled out because there are no 7-dimensional homogeneous manifolds that admit a transitive and effective
action of the $\mathfrak{t}_0$ subalgebra of the expected symmetry superalgebra of such backgrounds.  So we shall begin with backgrounds with $N_R=2$.  The homogeneous spaces are as those
in IIA and are given in (\ref{cnr2}).

The homogeneous space $S^7=\mathrm{Spin}(8)/\mathrm{Spin}(7)$ can be ruled out immediately.  This symmetric space does not admit invariant 2- and 3-forms.  Therefore $Y=H=0$.
Then a field equation in (\ref{iibf}) implies that $X=0$ as well and so the warp factor field equation in (\ref{iibeinst}) cannot be satisfied.

Similarly $S^7=\mathrm{Spin}(7)/G_2$ can also be ruled out as it does not admit an invariant closed 3-form and so $H=0$. Also  it does not admit an invariant 2-form either
and so $Y=0$.  Then because of the field equations in (\ref{iibf}), one deduces that $X=0$ and so the warp factor field equation in (\ref{iibeinst}) becomes inconsistent.

\subsubsection{$S^7=U(4)/U(3)$}

Following the description  for the geometry of the homogeneous space $U(4)/U(3)$ as in section \ref{s7u4u3}, the most general allowed fluxes are
\bea
Y=\alpha\, \omega~,~~~H=\beta\, \ell^7\wedge \omega~.
\eea
The Bianchi identity $dH=0$ requires that $\beta=0$. In turn a field equation in (\ref{iibf}) implies that $X=0$.  Substituting this back into the Bianchi identities
(\ref{iibb}), one finds that $Y$ is harmonic and so it must vanish.  As all fluxes vanish, the warp factor field equation in (\ref{iibeinst}) becomes inconsistent. There are no AdS$_3$ solutions with
internal space $U(4)/U(3)$.

\subsubsection{$S^7=(Sp(2)\times Sp(1))/Sp(1)\times Sp(1)$}

The geometry of this homogeneous space described in section \ref{sp2sp1sp1sp1s1} reveals that there are no invariant 2-forms and closed 3-forms.  As a result $Y=H=0$.
The field equations (\ref{iibf}) imply that $X=0$ as well.  Therefore there are no solutions as the warp factor field equation cannot be satisfied.

\subsubsection{$S^4\times S^3=\mathrm{Spin}(5)/\mathrm{Spin}(4)\times SU(2)$} \label{s533}

The geometry of this homogeneous space space has been described in section \ref{o5o4su2}.  The most general fluxes can be chosen as
\bea
Y={1\over2} \alpha_r \epsilon^r{}_{st} \ell^s\wedge \ell^t~,~~~H=\beta\, \ell^{123}~,
\eea
where $r,s,t=1,2,3$.
As $Y$ is both closed and co-closed and $H^2(S^4\times S^3)=0$, we deduce that $Y=0$ and  so a Bianchi identity in (\ref{iibb}) implies that
\bea
\bar X \beta- X \bar \beta=0~.
\eea
This together with  a field equation in (\ref{iibf}) imply $|\beta|^2+|X|^2=0$ and so $X=H=0$.
Then the warp factor field equation in (\ref{iibeinst}) cannot be satisfied. There are no AdS$_3$ solutions with internal space $\mathrm{Spin}(5)/\mathrm{Spin}(4)\times SU(2)$.

\subsection{$N>16$ solutions with $N_R=4$}

The homogeneous internal spaces are given in (\ref{cnr4}).  It is straightforward to show that $S^6\times S^1=\mathrm{Spin}(7)\mathrm{Spin}(6)\times S^1$ is not a solution
as $Y=H=X=0$ which contradicts the warp factor field equation.

\subsubsection{$G_2/SU(3)\times S^1$}

In the notation of section \ref{g2su3s1} the metric can be chosen as $ds^2=a \delta_{r\bar s} \lambda^r \lambda^{\bar s}+ b (\bbl^7)^2$ and the most general
invariant $Y$ and $H$ forms are
\bea
Y=\alpha\, \omega~,~~~H=\beta_1\, \mathrm{Im}\chi+ \beta_2\, \ell^7\wedge \omega+\beta_3\, \mathrm{Re}\chi~.
\eea
The Bianchi identity $dH=0$ implies that $\beta_2=\beta_3=0$. In what follows set $\beta_1=\beta$.  It follows that the Bianchi identity for $Y$ in (\ref{iibb})
gives
\bea
3\alpha={i\over8} (\bar X \beta- \bar\beta X)~.
\label{axxbb}
\eea
Furthermore the field equation for $H$ in (\ref{iibf}) gives
\bea
\beta=-i X\, a\, \alpha~.
\label{axxbb2}
\eea
Next turn to the dilatino KSE. Setting $\lambda^r=\ell^{2r-1}+i \ell^{2r}$, it can be written as
\bea
{\beta\over a^{{3\over2}}}(J_1+J_2-J_3-J_1J_2J_3) \sigma_+=X \sigma_+
\eea
where $J_1=\Gamma_z \Gamma_{136}$, $J_2=\Gamma_z \Gamma_{235}$ and $J_3=\Gamma_z \Gamma_{246}$.  These are commuting hermitian Clifford algebra operators with eigenvalues
$\pm 1$.  For all choices of eigenspaces either $X=0$ or $X=\pm 4\beta/a^{{3\over2}}$.  Substituting this into the first field equation in (\ref{iibf}), we find that $\beta=0$.
Therefore $X=0$ as well. Then (\ref{axxbb}) and (\ref{axxbb2}) imply that $Y=H=0$.  Thus  the warp factor field equation in (\ref{iibeinst})  cannot be satisfied.  Therefore there are no supersymmetric AdS$_3$ solutions with internal space
$G_2/SU(3)\times S^1$.

\subsection{$N>16$ solutions with $N_R=6$}

The allowed homogeneous internal spaces are given in (\ref{cnr6}).  We have already investigated the AdS$_3$ backgrounds with internal space $S^4\times S^3=\mathrm{Spin}(5)/\mathrm{Spin}(4)\times (SU(2)\times SU(2))/SU(2)$ as they are a special case of those explored in section \ref{s533} and  we have found that there are no solutions. Next we shall  examine the remaining two cases.

\subsubsection{$S^5\times S^2=\mathrm{Spin}(6)/\mathrm{Spin}(5)\times SU(2)/U(1)$}
The metric can be chosen as
\bea
ds^2(M^7)=  ds^2(S^5)+  ds^2(S^2)=a\, \delta_{rs} \ell^r \ell^s+ b\,\big((\ell^6)^2+(\ell^7)^2\big)~,
\eea
where $a,b>0$ are constants, $\ell^r$, $r=1,\dots, 5$ is a left-invariant frame on $S^5$ and $(\ell^6, \ell^7)$ is a left-invariant frame on $S^2$.
As this symmetric space does not admit  invariant 3-forms, we have  $H=0$.  Then a field equation in (\ref{iibf}) implies that $X=0$.  Setting $Y=\alpha\,\ell^{67}$,
the Einstein equation along $S^2$ gives
\bea
R_{pq}^{(7)}=-4{\alpha^2\over b^2} \delta_{pq}~,~~~p,q=6,7~.
\eea
However the Ricci tensor of $S^2$ is strictly positive.  Thus there are no AdS$_3$ solutions with internal space $\mathrm{Spin}(6)/\mathrm{Spin}(5)\times SU(2)/U(1)$.


\subsubsection{$S^5\times S^2=U(3)/U(2)\times SU(2)/U(1)$}

The geometry of this homogeneous space has already been described in section \ref{u3u2su2u1} and the metric is given in (\ref{metru3u2su2u1}).
The most general fluxes can be chosen as
\bea
Y=\alpha_1\,\omega+ \alpha_2\, \sigma~,~~~H=\beta_1\, \ell^5 \wedge \omega+\beta_2\, \ell^5\wedge \sigma~.
\eea
The Bianchi identity $dH=0$ implies that $\beta_1=\beta_2=0$ and so $H=0$.  In turn a field equation in (\ref{iibf})  gives that $X=0$, and $d*_7Y=0$ implies $\alpha_1=0$.

Next consider the Einstein equation (\ref{iibeinst}) along $S^2$.  A direct calculation reveals that
\bea
R_{pq}^{(7)}=-4{\alpha_2^2\over b^2}\, \delta_{pq}~,~~~p,q=6,7~.
\label{s2ein}
\eea
However the Ricci tensor of $S^2$ is strictly positive.  There are no AdS$_3$ solutions with internal space $U(3)/U(2)\times SU(2)/U(1)$.

\subsection{$N>16$ solutions with $N_R=8$}

The allowed homogeneous internal spaces are given in (\ref{cnr8}).  All the cases have already been investigated apart from those with internal space
$S^4\times S^2 \times S^1=\mathrm{Spin}(5)/\mathrm{Spin}(4)\times SU(2)/U(1)\times S^1$ and $N^{k,l,m}$ which we shall  examine next.

\subsubsection{$S^4\times S^2 \times S^1=\mathrm{Spin}(5)/\mathrm{Spin}(4)\times SU(2)/U(1)\times S^1$}

The geometry of this symmetric space has been described in section \ref{so5so4su2u1s1}. The metric can be chosen as
in (\ref{metrso5so4su2u1s1}) and the most general invariant fluxes are
\bea
Y=\alpha\, \ell^{56}~,~~~H=\beta\, \ell^{567}~.
\eea
As $dY=0$, the Bianchi identities (\ref{iibb}) give that
\bea
\bar X \beta- \bar\beta X=0~,
\eea
which together with a field equations in (\ref{iibf}) imply that $H=X=0$.  The only non-vanishing field is $Y$.  However as in the previous case after evaluating the
Einstein equation along $S^2$, one finds a similar relation to (\ref{s2ein}).  This is a contradiction as the Ricci tensor of $S^2$ is strictly positive and so there are
no AdS$_3$ solutions with internal space $\mathrm{Spin}(5)/\mathrm{Spin}(4)\times SU(2)/U(1)\times S^1$.

\subsubsection{$N^{k,l,m}=(SU(2)\times SU(3)\times U(1))/\Delta_{k,l,m} (U(1)\times U(1))\cdot (1\times SU(2))$}

The metric can be chosen as in (\ref{metrnklm}).
 From the results of appendix \ref{appencx2}, one can deduce that there
are no closed invariant 3-forms and so $H=0$. The field equations (\ref{iibf}) imply that $X=0$ as well. The most general 2-form $Y$ is
\bea
Y=\alpha_1\, \omega_1+ \alpha_2\, \omega_2~.
\eea
The Bianchi identities imply that $Y$ must be harmonic.  Observe that $dY=0$.  The co-closure condition implies that
\bea
{\alpha_1 \over l\, b^2}-{\alpha_2 \over k\, c^2} =0~,
\label{coco}
\eea
where $d\text{vol}={1\over2}\omega_1^2\wedge \omega_2\wedge \ell^7$.

Next the algebraic KSE (\ref{algkseiib}) can be written as
\bea
\left({\alpha_1\over b} (J_1+J_2)+{\alpha_2\over c} J_3\right)\sigma_+={1\over\ell A} \sigma_+~,
\label{nkij}
\eea
where $J_1=i\Gamma^{12}$, $J_2=i \Gamma^{34}$ and $J_3=i\Gamma^{56}$. The relations amongst the fluxes for each of the eigenspaces can be found in table \ref{tableklx}.  The warp factor field equation in (\ref{iibeinst}) also gives
\bea
4{\alpha_1^2\over b^2}+2 {\alpha_2^2\over c^2}={1\over \ell^2 A^2}~.
\eea
As the common eigenspaces of $J_1, J_2, J_3$ have dimension 2   to find solutions preserving $N>16$ supersymmetries, one needs to choose at least three such  eigenspaces.

\begin{table}[h]
\begin{center}
\vskip 0.3cm
 \caption{Decomposition of (\ref{nkij})  into eigenspaces}
 \vskip 0.3cm

	\begin{tabular}{|c|c|}
		\hline
		$|J_1,J_2, J_3\rangle$&  relations for the fluxes\\
		\hline
		$|\pm,\mp,+\rangle$& ${\alpha_2\over c} ={1\over \ell A} $ \\
\hline
		$|\pm,\mp,-\rangle$& ${\alpha_2\over c} =-{1\over \ell A} $ \\
		\hline
		$|\pm,\pm,\pm\rangle$& ${2\alpha_1\over b}+{\alpha_2\over c} =\pm{1\over \ell A} $ \\
		\hline
$|\pm,\pm,\mp\rangle$& ${2\alpha_1\over b}-{\alpha_2\over c} =\pm{1\over \ell A} $ \\
		\hline
		\end{tabular}
\vskip 0.2cm
  \label{tableklx}
 \end{center}
\end{table}

The eigenspaces that lead to the relation ${\alpha_2\over c} =\pm{1\over \ell A} $  for the fluxes can be ruled out because of the warp factor field equation.  Therefore we have to choose
three eigenspaces from the remaining cases in table \ref{tableklx}.  For every choice of a pair of relations either $\alpha_1$ or $\alpha_2$ vanishes. Then the co-closure condition (\ref{coco}) implies that
$Y=0$. There are no AdS$_3$ supersymmetric solutions preserving $N>16$ supersymmetries.

\section*{Acknowledgments}

One of us GP thanks Hermann Nicolai for an invitation to visit the Max Planck Institute for Gravitational Physics (Albert Einstein   Institute) in Golm  where part of this work has been carried out.
The work of ASH is supported by the German Science Foundation (DFG) under the Collaborative Research Center (SFB) 676 ``Particles, Strings and the Early Universe''. GP is partially supported from the  STFC rolling grant ST/J002798/1.

\setcounter{section}{0}\setcounter{equation}{0}

 \begin{appendices}
\appendix{Notation and conventions}

Our  conventions for forms are as follows. Let $\omega$ be a k-form, then
\bea
\omega=\frac{1}{k!} \omega_{i_1\dots i_k} dx^{i_1}\wedge\dots \wedge dx^{i_k}~,~~~\omega^2_{ij}= \omega_{i\ell_1\dots \ell_{k-1}} \omega_{j}{}^{\ell_1\dots \ell_{k-1}}~,~~~
\omega^2= \omega_{i_1\dots i_k} \omega^{i_1\dots i_k}~.
\eea
To simplify expressions, we use the shorthand notation
\bea
\omega^n=\omega\wedge\dots\wedge\omega~,~~~\ell^{12\dots n}=\ell^1\wedge \ell^2\wedge\dots\wedge \ell^n
\eea
where $\ell^i$ is a left-invariant frame, and similarly for an orthonormal frame $\bbe^i$.

Our Hodge duality convention is
\bea
\omega\wedge *\omega={1\over k!} \omega^2 d\mathrm{vol}~.
\eea
We also define
\bea
{\slashed\omega}=\omega_{i_1\dots i_k} \Gamma^{i_1\dots i_k}~, ~~{\slashed\omega}_{i_1}= \omega_{i_1 i_2 \dots i_k} \Gamma^{i_2\dots i_k}~,~~~\slashed{\gom}_{i_1}= \Gamma_{i_1}{}^{i_2\dots i_{k+1}} \omega_{i_2\dots i_{k+1}}~,
\eea
where the $\Gamma_i$ are the Dirac gamma matrices.  Throughout the paper, the gamma matrices are always taken in an orthonormal frame.

The inner product $\langle\cdot, \cdot\rangle$ we use on the space of spinors is that for which space-like gamma matrices are hermitian while time-like gamma
matrices are anti-hermitian, i.e. the Dirac spin-invariant inner product is $\langle\Gamma_0\cdot, \cdot\rangle$.  For more details on our conventions
see \cite{mads, iibads, iiaads}.

\appendix{Coset spaces  with  non-semisimple transitive groups }\label{modif}

We have already argued that if $\mathfrak{t}_0$ is simple then the internal spaces $G/H$, with $\mathfrak{Lie}\,G=\mathfrak{t}_0$ can be identified from the classification results of
\cite{castellani}-\cite{bohmkerr}.  This is also the case for $\mathfrak{t}_0$ semisimple provided that in addition one considers modifications to the homogeneous spaces as described in section \ref{modifx}.

Here, we shall describe the structure of homogeneous $G/H$ spaces for  $G$ a compact  but  not semisimple Lie group.  As  we consider homogeneous spaces   up to discrete identifications, we shall perform
the calculation in terms of Lie algebras.
 The Lie algebra of $G$ can be written as $\mathfrak{Lie}\,G=\mathfrak{p}\oplus \mathfrak{a}$, where $\mathfrak{p}$
is semisimple and $\mathfrak{a}$ is abelian.  Suppose now that we have a $G/H$ coset space, where $H$ is a compact subgroup of $G$. Then $\mathfrak{Lie}\,(H)=\mathfrak{q}\oplus \mathfrak{b}$, where $\mathfrak{q}$ is a semisimple subalgebra
and $\mathfrak{b}$ is abelian.  Let us now focus on the inclusion $i~:~~\mathfrak{Lie}\,(H)\rightarrow \mathfrak{Lie}\,G$.  Consider the projections $p_1~:~~\mathfrak{Lie}\,G\rightarrow \mathfrak{p}$ and $p_2~:~~\mathfrak{Lie}\,G\rightarrow \mathfrak{a}$.  Then we have that $p_2\circ i\vert_{\mathfrak{q}}=0$ as there are no non-trivial Lie algebra homomorphisms from
a semisimple Lie algebra into an abelian one.  Thus $p_1\circ i\vert_{\mathfrak{q}}$ is an inclusion.  Furthermore $p_1\circ i\vert_{\text{Ker} (p_2\circ i\vert_{\mathfrak{b}})}$
is also an inclusion.  Therefore $\mathfrak{q}\oplus \text{Ker} (p_2\circ i\vert_{\mathfrak{b}})$ is a subalgebra of $\mathfrak{p}$.  As $\mathfrak{p}$ is semisimple
and for applications here $\text{dim}\, G/H\leq 8$, there is a classification of all coset spaces $P/T$ with $\mathfrak{Lie}\, P=\mathfrak{p}$  and
$\mathfrak{Lie}\,(T)=\mathfrak{q}\oplus \text{Ker}\, p_2\circ i\vert_{\mathfrak{b}}$.

Next consider $\mathfrak{b_1}=\mathfrak{b}/\text{Ker}\, (p_2\circ i\vert_{\mathfrak{b}})$.  Suppose first that $i(\mathfrak{b}_1)$ is contained in both $\mathfrak{p}$ and $\mathfrak{a}$, then up to a discrete identification a coset space $W/X$ with $\mathfrak{Lie}\,W=\mathfrak{p}\oplus i(\mathfrak{b})$ and $\mathfrak{Lie}\,X=\mathfrak {q}\oplus \mathfrak{b}$ is a modification of $P/T$ with an abelian group which has Lie algebra $p_2\circ i(\mathfrak{b})$.  Furthermore the generators
of  $\mathfrak{a}/i(\mathfrak{b})$ commute with $\mathfrak{Lie}\,X$ and of course are not in the image of $i$.  As a result $G/H$ up to a discrete identification
can be written as $W/X\times T^k$, where $k=\text{dim}\, \mathfrak{a}/i(\mathfrak{b})$, ie up to a discrete identification $G/H$ is the product of an abelian  modification of a coset space with semisimple
transitive group and of an abelian group. On the other hand if $i(\mathfrak{b})$ is all contained in $\mathfrak{a}$, then $G/H$ up to a discrete identification is a product
$P/T\times T^m$, where $m=\text{dim}\,i(\mathfrak {a})$.
In the classification of AdS$_3$ backgrounds, we  use the above results to describe the geometry of the internal spaces
whenever $\mathfrak{t}_0$ is not a semisimple Lie algebra.

\appendix{{\large $N^{k,l}=SU(2)\times SU(3)/ \Delta_{k,l}(U(1))\cdot (1\times SU(2))$}}\label{appencx2}

The inclusion of $U(1)\times SU(2)$ in $SU(2)\times SU(3)$ is given by
$$
(z, A)\rightarrow \Big( \begin{pmatrix} z^k & 0\\ 0 & z^{-k} \end{pmatrix}, \begin{pmatrix} A z^l &0\\ 0 & z^{-2l} \end{pmatrix}\Big)~.
$$
Consequently in the notation of \cite{ads4Ngr16}, the Lie subalgebra $\mathfrak{h}$ of the isotropy group is identified as
\bea
\mathfrak{h}=\mathbb{R}\langle  M_{12},  N_{12}, N_{11}+{1\over2} N_{33}, 2k \tilde N_{11}-3 l N_{33}\rangle~.
\eea
The generators of the tangent space at the origin $\mathfrak{m}$  of the homogeneous space must be linearly independent from those of $\mathfrak{h}$
and so one can choose
\bea
\mathfrak{m}=\mathbb{R}\langle \tilde M_{12}, \tilde N_{12}, M_{13}, M_{23}, N_{13}, N_{23}, 2k \tilde N_{11}+ 3\l N_{33}\rangle~,
\eea
where $\mathfrak{su}(2)=\mathbb{R}\langle \tilde M_{rs}, \tilde N_{rs}\rangle$ , $r,s=1,2$ and $\mathfrak{su}(3)=\mathbb{R}\langle  M_{ab},  N_{ab}\rangle$ , $a,b=1,2,3$.

A left-invariant frame on $N^{k,l}$ is
\bea
\ell=\ell^7 Z+ \ell^5 \tilde M_{12}+\ell^6 \tilde N_{12} + \ell^r M_{r3}+ \hat \ell^r N_{r3}~,
\eea
where $Z=2k \tilde N_{11}+ 3\l N_{33}$.

The exterior differential algebra of the left-invariant frame, modulo the terms  that contain the canonical connection and so lie in $\mathfrak{h}\wedge \mathfrak{m}$,  is
\bea
d\ell^7&=&-{1\over 4k}\ell^5\wedge \ell^6+ {1\over 8l} \delta_{rs} \ell^r\wedge \hat \ell^s~,
\cr
d\ell^5&=&2k\ell^7\wedge \ell^6~,~~~~d\ell^6=-2k\ell^7\wedge\ell^5~,
\cr
d\ell^r&=&-3l\,\ell^7\wedge \hat\ell^r~,~~~d\hat \ell^r=3l\,\ell^7\wedge \ell^r~.
\eea
Note that upon taking the exterior derivative of invariant forms the terms in the exterior derivative of a left-invariant frame that lie in $\mathfrak{h}\wedge \mathfrak{m}$ do not contribute.

The invariant forms  on $N^{k,l}$ are generated by a 1-form $\ell^7$ and the 2-forms
\bea
\omega_1= \delta_{rs} \ell^r\wedge \hat\ell^s~,~~~\omega_2=\ell^5\wedge \ell^6~.
\eea
 Observe that $d\omega_1=d\omega_2=0$.  On the other hand $d\ell^7=-(4k)^{-1} \omega_2+ (8l)^{-1} \omega_1$.  So $H^2(M^7, \bR)$ has one generator as expected.
The invariant 3- and  4-forms are $\ell^7\wedge \omega_1$, $\ell^7\wedge \omega_2$,  and $\omega_1\wedge \omega_1$,  $\omega_1\wedge \omega_2$, respectively.  Both 4-forms are exact as they are the exterior derivatives of  invariant 3-forms.  As a
result $H^4(M^7, \bR)=0$ as expected.  Note though that $H^4(M^7, \bZ)\not=0$.

\end{appendices}

\end{document}